\documentclass{article}

\usepackage[utf8]{inputenc}
\usepackage{float}
\usepackage{jheppub}
\usepackage{amsmath}
\usepackage[english]{babel}
\usepackage{amssymb}
\usepackage{mathtools}
\usepackage{color}
\definecolor{ao}{rgb}{0.0, 0.5, 0.0}
\newcommand{\red}[1]{\textcolor{black}{#1}}

\newcommand{\avg}[1]{\langle #1 \rangle}

\usepackage{graphicx}
\usepackage{longtable}
\usepackage{tikz}
\usetikzlibrary{backgrounds}
\usepackage{comment}
\usepackage{capt-of}
\usepackage{amsthm}
\usepackage{hyperref}
\hypersetup{
    colorlinks=true,
    linkcolor=blue
    }

\usepackage{bm}

\usepackage{caption}
\usepackage{subcaption}

\newcommand{\mycomment}[1]{}

\def\0{\mbox{\tiny $0$}}
\def\1{\mbox{\tiny $1$}}
\def\2{\mbox{\tiny $2$}}
\def\3{\mbox{\tiny $3$}}
\def\4{\mbox{\tiny $4$}}
\def\5{\mbox{\tiny $5$}}
\def\6{\mbox{\tiny $6$}}
\def\7{\mbox{\tiny $7$}}
\def\8{\mbox{\tiny $8$}}
\def\9{\mbox{\tiny $9$}}

\usepackage{amsmath} % for aligned
\usepackage{tikz}
\usepackage{listofitems} % for \readlist to create arrays
\usetikzlibrary{arrows.meta} % for arrow size
\usepackage[outline]{contour} % glow around text
\contourlength{1.4pt}

\usepackage[ruled,vlined]{algorithm2e} % Per gli algoritmi

\tikzset{>=latex} % for LaTeX arrow head
\usepackage{xcolor}
\colorlet{myred}{red!80!black}
\colorlet{myblue}{blue!80!black}
\colorlet{mygreen}{green!60!black}
\colorlet{myyellow}{yellow!60!black}
\colorlet{myorange}{orange!85!red!90!black}
\colorlet{mydarkred}{red!30!black}
\colorlet{mydarkblue}{blue!40!black}
\colorlet{mydarkgreen}{green!30!black}
\tikzstyle{node}=[thick,circle,draw=myblue,minimum size=22,inner sep=0.5,outer sep=0.6]
\tikzstyle{node1}=[thick,rectangle,draw=myblue,minimum width = 0.7cm,  minimum height = 0.7cm,inner sep=-0.01,outer sep=0.01]
\tikzstyle{node in}=[node,green!20!black,draw=mygreen!30!black,fill=mygreen!25]
\tikzstyle{nodesigma}=[node,green!20!black,draw=mygreen!30!black,fill=mygreen!10]
\tikzstyle{nodetau}=[node,orange!20!black,draw=mygreen!30!black,fill=myorange!20]
\tikzstyle{nodephi}=[node,green!20!black,draw=mygreen!30!black,fill=myred!20]
\tikzstyle{node hidden}=[node1,blue!20!black,draw=myblue!30!black,fill=myblue!05]
\tikzstyle{node convol}=[node,orange!20!black,draw=myorange!30!black,fill=myorange!20]
\tikzstyle{node out}=[node,red!20!black,draw=myred!30!black,fill=myred!20]
\tikzstyle{connect}=[thick,mydarkblue] %,line cap=round
\tikzstyle{connect arrow}=[-{Latex[length=4,width=3.5]},thick,mydarkblue,shorten <=0.5,shorten >=1]
\tikzset{ % node styles, numbered for easy mapping with \nstyle
  node 1/.style={node in},
  node 2/.style={node hidden},
  node 3/.style={node out},
}
\usetikzlibrary{fit}
\newcommand{\lr}[1]{\left(#1\right)}

\newcommand{\SOMMA}[2]{\displaystyle\sum\limits_{#1}^{#2}}

\long\def \beq#1\eeq {\begin{equation} #1 \end{equation}}
\long\def \beaq#1\eeaq {\begin{equation}\begin{aligned} #1 \end{aligned}\end{equation}}
\long\def \bes#1\ees {\begin{equation}\begin{split} #1 \end{split} \end{equation}}
\long\def \bea#1\eea {\begin{eqnarray} #1 \end{eqnarray}}
\long\def \bse[#1]#2\ese {\begin{subequations}\label{#1}\begin{align} #2 \end{align}\end{subequations}}

\title{Networks of \red{Hebbian} networks: {\em more is different}}

\author[a,b]{Elena Agliari,}
\author[c,d]{Andrea Alessandrelli,}
\author[b,e]{Adriano Barra,}
\author[f]{Martino Salomone Centonze,}
\author[g,h,i]{Federico Ricci-Tersenghi}

\affiliation[a]{Dipartimento di Matematica, Sapienza Universit\`a di Roma, Rome, Italy.}

\affiliation[b]{Istituto Nazionale d'Alta Matematica, GNFM, Roma, Italy.}

\affiliation[c]{Dipartimento di Informatica, Università di Pisa, Pisa Italy.}

\affiliation[d]{Istituto Nazionale di Fisica Nucleare, Sezione di Lecce, Italy.}

\affiliation[e]{Dipartimento di Scienze di Base Applicate all'Ingegneria, Sapienza Universit\`a di Roma, Rome, Italy.}

\affiliation[f]{Dipartimento di Matematica, Universit\`a di Bologna, Italy.}

\affiliation[g]{Dipartimento di Fisica, Sapienza Universit\`a di Roma, Rome, Italy.}

\affiliation[h]{Istituto Nazionale di Fisica Nucleare, Sezione di Roma1, Italy.}

\affiliation[h]{CNR-Nanotec, Rome unit, 00185 Roma, Italy.}

\abstract{The common thread behind the recent Nobel Prize in Physics to John Hopfield and those conferred to Giorgio Parisi in 2021 and Philip Anderson in 1977 is \emph{disorder}.  Quoting Philip Anderson: \emph{more is different}. This principle has been extensively demonstrated in magnetic systems and spin glasses, and, in this work, we test its validity on Hopfield neural networks to show how an assembly of these models displays emergent capabilities that are not present at a single network level.  Such an assembly is designed
as a layered associative Hebbian network that, beyond accomplishing standard \emph{pattern recognition},
spontaneously performs also \emph{pattern disentanglement}. Namely, when inputted with a composite signal – e.g., a musical chord – it can return the single constituting elements – e.g., the notes making up the chord. Here, restricting to notes coded as Rademacher vectors and chords that are their mixtures (i.e., spurious states), we use tools borrowed from statistical mechanics of disordered systems  to  investigate this task, obtaining the conditions
over the model control-parameters such that pattern disentanglement is successfully executed.}

\begin{document}

\maketitle

\section{Introduction}

The celebrated constructive criticism to the reductionist hypothesis {\em more is different} -- a concept popularized by Philip W. Anderson in the 70's \cite{AndersonScience} -- is a foundational statement\footnote{\red{The phrase emphasizes that, in a complex network, collective phenomena emerge that cannot be predicted from -- or reduced to -- the behavior of its isolated nodes: for instance, the Hopfield model is able to recognize a pattern due to the inner dialogues among its neurons (i.e. the nodes of the network) but none of them has even the concept of pattern, thus  the question addressed in this manuscript: are new collective phenomena appearing by constructing networks of Hopfield models?}} in Statistical Mechanics and its manifestations are ubiquitous in Nature, from phase transitions in Physics \cite{Derrida1984,Goldenfeld2018} and Chemistry \cite{FrenkelSmith,EyesChem} to collective behaviors in Biology \cite{CavagnaCabibbo,Bialek2012} and Ecology \cite{Sole2006,Chiara2021}. In this paper, we inspect this principle at work with Hopfield associative neural networks \cite{Hopfield}, each of which, independently, can perform only a specific task, that is, {\em pattern recognition} \cite{Bishop}. % and play as {\em distributed associative memory}. 
\newline
In particular, we consider an ensemble of Hopfield networks that share the same dataset of random, binary patterns \cite{Coolen} and couple them through repulsive interactions. Our findings demonstrate that the resulting network of networks can execute tasks that exceed the capabilities of any single constituting network. 
Specifically, the combined system exhibits the ability to perform {\em pattern disentanglement} —i.e., when presented with a mixture of patterns, it can separate the input into the original components.  In fact, a composite system of, say, $L$ Hopfield networks displays the natural architecture to disentangle combinations of $L$ patterns; the mixtures that we will consider here are obtained by applying a majority rule to $L$ patterns drawn from the dataset: this produces the so-called {\em spurious states}, known to emerge as (unwanted) minima in a single Hopfield model \cite{Amit}. 
\newline
It is worth noticing that our assembly of $L$ interacting Hopfield networks can also be looked at as an $L$-directional associative memory \cite{Kosko,BCDS-JPA2023,AMT-JSP2018} endowed with Hebbian interactions where intra-layer interactions are attractive but inter-layer interactions are repulsive (i.e. their sign is reversed, unlike classic directional associative memories). Without such a reversal, pattern disentanglement would be prevented as layers would tend to align to the same pattern, unless the task is simplified to disentangling mixtures of patterns drawn from independent datasets (a simpler task\footnote{In that scenario, mixtures states can not be seen as Hopfield spurious states.} that can be handled by standard hetero-associative neural networks \cite{AABCR-2024}).
\par\medskip
From a theoretical standpoint, this new capability of the model under study allows for further dissecting the world of spurious states and it may shed further light on the complex landscape of the Hopfield model itself. On the practical side, the potential applications are vast. Recalling that the most stable spurious states of the Hopfield model are mixtures built of by triplets of patterns \cite{Amit}, one can consider, for instance, video signals, where colors emerge from the combination of three primary colors (red, yellow, and blue), or audio signals, where chords consist of three primary notes (as, e.g. the C-Major chord is a triad formed from a root C, a major third E and a perfect fifth G). However, rather than focusing on specific applications, our aim here is to construct a quantitative theory able to describe the network’s emergent computational properties and uncover the fundamental mechanisms underpinning them, in the context of synthetic datasets.
%%%%%%%%%%%%%%%%%%%%%%%%%%%%%%%%%%%%%%%%%%%%
\par\medskip
\red{It is worth pointing out that, at present, there are already several algorithmic approaches to pattern disentanglement, yet nor any of these is based on Hebbian learning, neither any of these provides a theoretical explanation for this type of information processing by neural networks. For instance, \emph{Disentangled Representation Learning} \cite{DRL2024} plays a central role when the network fails to infer the correct features (as often they have to be disentangled at first) and this, in turn, turned pivotal for explainable AI in visual recognition  (e.g., when an image contains several objects to be recognized). Typically, the underlying neural architectures are deep learning scaffolds (namely long multi-layered networks or deep Boltzmann machines) \cite{Cacao20} where, as the disentanglement goes by, inner layers naturally split into sub-architectures specialized in the recognition of a specific feature or pattern: \emph{Deep Hierarchical Representations} \cite{DipLer16} fall in this ensemble  too.}
\par \medskip
The paper is structured as follows. In Sec.~\ref{sec:model} we introduce the model \red{and Sec.~\ref{sec:model2}} we present the main analytical results obtained by employing statistical-mechanics tools. Next, \red{focusing on the test-case $L=3$,} in Sec.~\ref{sec:appl1}, we explore its ability to perform pattern disentanglement by different approaches. In Sec.~\ref{sec:HLN}, we study analytically the stability of several paradigmatic configurations, e.g., where each layer is aligned with the $L$-pattern mixtures and where each layer is aligned with a different pattern participating in the mixture: these two configurations play, respectively, as the network input and the target network output. Next, in Sec.~\ref{sec:LLNoisy}, we make this analysis more accurate by examining the sign of the free-energy Hessian matrix, whence we get insights on the stability of the input and of the target configurations; then, in Sec.~\ref{sec:ssr}, we proceed the investigation by finding numerical solutions of the self-consistency equations stemming from the statistical-mechanics analysis and by suitably revising the standard protocols designed to check retrieval capabilities in order to account for the disentanglement task; finally, in Sec.~\ref{sec:MC}, the previous theoretically-driven results are corroborated by \red{Monte Carlo (MC)} simulations. In the final Sec.~\ref{sec:concl}, we summarize results and discuss some outlooks. Technical details on analytical computations are collected in the Appendices \ref{Appendix-Guerra}-\ref{sec:spectrum}. \red{Moreover, in Appendix \ref{ssec:MC_details} we describe the methodology underlying computational experiments and in Appendix \ref{sec:L5} we check the robustness of the results by running analogous experiments, but setting $L=5$.} Next, possible adjustments to the model that could enhance its performance are discussed in Appendix \ref{sec:NewModel}. Finally, in Appendix H, we show how our theory for pattern disentanglement by the present Hebbian network can also shed light on pattern disentanglement by already existing architectures.

%\section{$L$-directional associative memory} \label{sec:model}
%\section{Hebbian networks of networks} \label{sec:model}
\section{\red{Definitions}} \label{sec:model}
\begin{figure}
    \centering
    \includegraphics[width=8cm]{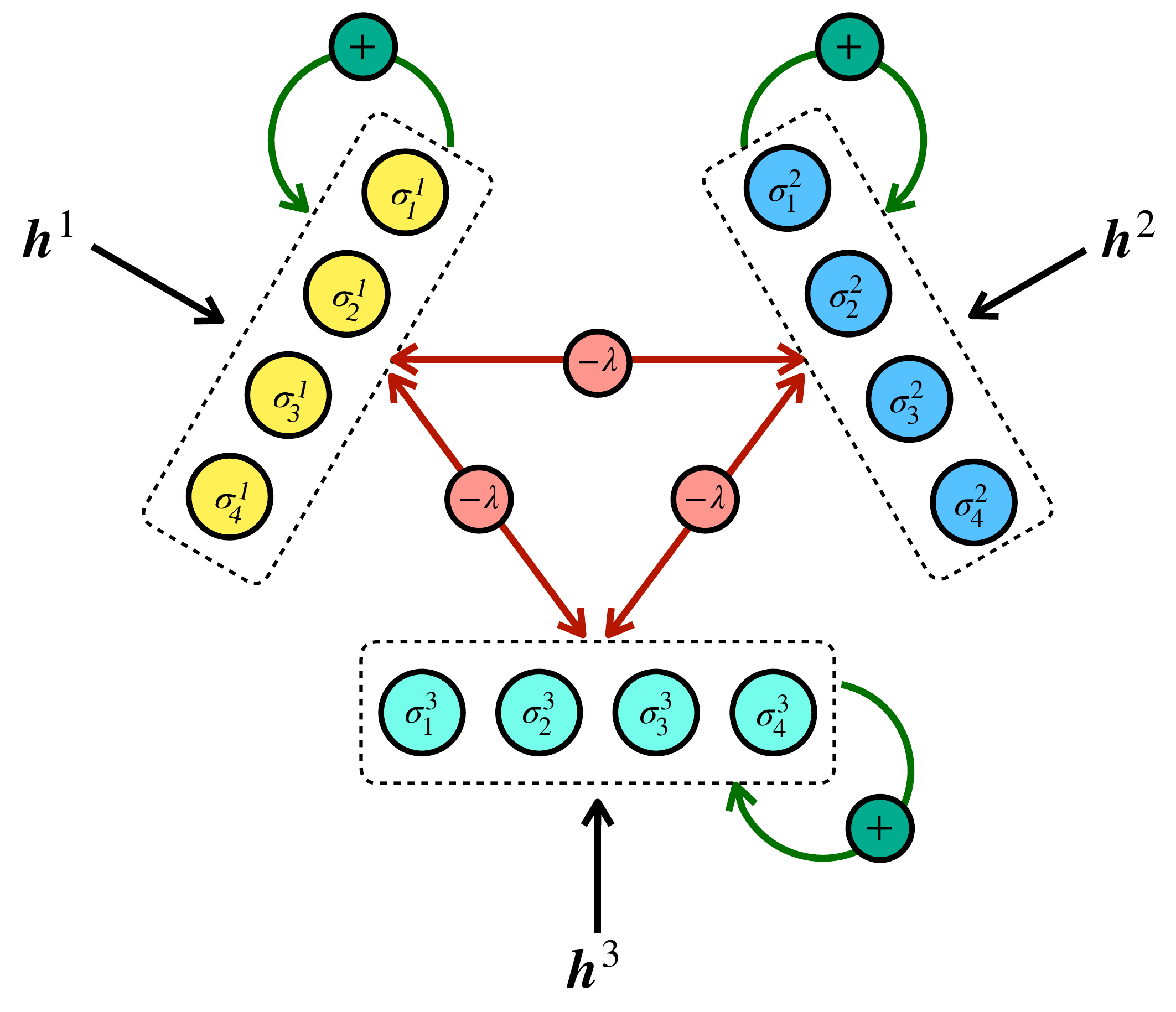}
    \caption{\red{Schematic representation of the model under study, \red{where we set $L=3$}. The three contributions making up the \red{cost function} \eqref{eq:HamHam} are highlighted: imitative intra-layer interactions (represented by a $\oplus$ loop), anti-imitative inter-layer interactions  (represented by a $-\lambda$ double arrow) and the coupling with an external field (represented by a single arrow \red{$\boldsymbol h$}).}}
    \label{fig:netw_rapp}
\end{figure}

In this section we introduce the general model, whose architecture is sketched in Figure~\ref{fig:netw_rapp}; to avoid ambiguities, we will refer to a single Hopfield network as a layer. Thus, let us consider $L$ layers, each composed of $N$ binary neurons, denoted as $\boldsymbol \sigma^a =(\sigma^a_1, ..., \sigma^a_N) \in \{-1, +1, \}^N$ for $a=1,...,L$, that interact pairwise as specified by the following cost function (or \emph{energy} or \emph{Hamiltonian}):
\begin{align} \label{eq:hamiltonian0}
   \mathcal  H(\boldsymbol \sigma ; \boldsymbol g, \boldsymbol \xi ) = %- \frac{1}{N} M^T G M = 
    -\frac{1}{N} \sum_{\mu=1}^K \sum_{a,b=1}^L g_{ab} \sum_{i,j=1}^N \sigma^a_i \xi^\mu_i \xi^\mu_j  \sigma^b_j,
    %= 
    %-N \sum_{\mu=1}^K \sum_{a,b=1}^L g_{ab} m^a_\mu m^b_\mu
\end{align}

where $\boldsymbol \xi^{\mu} =(\xi^{\mu}_1,...,\xi^{\mu}_N) \in \{-1, +1\}^N$ is the $\mu$-th pattern for $\mu=1,...,K$ and $\boldsymbol g \in \mathbb R^{L \times L}$ specifies the nature (imitative or anti-imitative) of the Hebbian intra-layer and inter-layer interactions.
By introducing the Mattis magnetizations
\begin{equation} \label{eq:Mattis}
    m^a_{\mu} = \frac{1}{N} \sum_{i=1}^N \xi_i^{\mu}\sigma_i^a, 
\end{equation}
as order parameters that assesses the retrieval of the $\mu$-th pattern by the $a$-th layer\footnote{\red{Note that ech $m_{\mu}^a \in [-1,+1]$ such that a value of $m_{\mu}^a$ close to $+1$ accounts for a retrieved pattern (likewise $m_{\mu}^a$ close to $-1$ accounts the retrieval of the inverse pattern $-\xi^{\mu}$) while $m_{\mu}^a \sim 0$ implies no retrieval.}}, we can recast \eqref{eq:hamiltonian0} as
\begin{align} \label{eq:hamiltonian}
   \mathcal  H(\boldsymbol \sigma ; \boldsymbol g, \boldsymbol \xi ) = -N \sum_{\mu=1}^K \sum_{a,b=1}^L  m^a_\mu g_{ab} m^b_\mu
\end{align}
thus, if $g_{ab} >0$ ($g_{ab} <0$), neurons tend to arrange in such a way that $\boldsymbol m^a \cdot \boldsymbol m^b$ is maximized (minimized). 
In the following we will restrict to this kind of structure:
\begin{equation} \label{eq:g_struct}
g_{ab}=
\begin{cases}
~~ 1 ~ \textrm{if}~ a=b\\
-\lambda ~ \textrm{if}~ a \neq b
\end{cases}
\end{equation}
with $\lambda \in [0, (L-1)^{-1})$ to ensure that $\boldsymbol g$ is positive definite (\emph{vide infra}). This implies that neurons belonging to the same layer interact by imitative Hebbian coupling -- namely, each layer tends to align to a single pattern, as it is the case in the standard Hopfield model -- while neurons belonging to different layers interact by anti-imitative Hebbian coupling -- namely, configurations where all layers are aligned with the very same pattern are discouraged, consistently with the kind of task we are interested in.  In any case, we stress that the Hebbian shape of the interaction is preserved and, as
%A sketch of the structure resulting from this choice is presented in Figure~\ref{fig:netw_rapp}.
%
%\begin{align}
%    g_{ab} = 
%    \begin{bmatrix}
%    1 & -\lambda & -\lambda\\
%    -\lambda & 1 & -\lambda\\
%    -\lambda & -\lambda & 1
%\end{bmatrix}
%\end{align}
%with $0\leq\lambda\leq\frac{1}{2}$ (the matrix is symmetric and positive %definite for $\lambda\leq\frac{1}{2}$). 
%
%
%As 
expected, in the limit $\lambda\to 0$ the model reduces to a collection of $L$ independent Hopfield models trained on the same dataset of patterns $\{\boldsymbol \xi^\mu\}_{\mu=1,..,K}$. 
The structure of the cost function \eqref{eq:hamiltonian} resembles that of $L$-directional associative memories \cite{Kosko,Peliti,BCDS-JPA2023,AABCR-2024}, but in those models intra-layer couplings are absent and the inter-layer couplings are imitative.
%namely, they discourage configurations where all layers are aligned with the very same pattern, consistently with the kind of task we are interested in.
\red{A modular organization of recurrent associative memories (but devoid of intra-layer interactions and exhibiting a hierarchical structure) has been proposed also in \cite{DmitryK2021}, inspired by cortical feedback structures, however that kind of network is most suitable for retrieving multiple objects within the same image and not for disentangling spurious mixtures as those faced here.}
\medskip

In general, we can allow for an external field, tuned by the scalar $H \in \mathbb R^+$ and pointing in the direction specified by $\boldsymbol h^a \in \{-1 , +1 \}^N$ for $a=1, ..., L$, namely
\begin{equation} \label{eq:HamHam}
\begin{array}{lll}
      \mathcal{H}(\boldsymbol \sigma ; \lambda, H, \boldsymbol \xi, \boldsymbol h ) = -N\SOMMA{\mu=1}{K} \sum_{a=1}^L (m^a_{\mu})^2 - H \SOMMA{i=1}{N}\sum_{a=1}^L h^{a}_i\sigma_i^a + N\dfrac{\lambda}{2}\SOMMA{\mu=1}{K} \sum_{\substack{a,b=1 \\ a\neq b}}^L m^a_\mu m^b_\mu . 
\end{array}
\end{equation}

Notice the variables and the parameters which the cost function depends on: beyond the model's degrees of freedom $\boldsymbol \sigma$, there appear the external fields $\boldsymbol h = \{\boldsymbol h^{a}\}_{a=1,...,L}$ that are quenched and will be chosen according to the application we aim to address with these networks\footnote{\red{We anticipate that, in this framework, we aim to assess the capacity of the model to tease apart the patterns appearing in mixtures like $\textrm{sign}(\sum_{\mu=1}^L \boldsymbol \xi^{\mu})$, as these are supplied as input. Thus, a natural choice for the external field is precisely $\boldsymbol h^{a} = \textrm{sign}(\sum_{\mu=1}^L \boldsymbol \xi^{\mu})$, for $a=1,...,M$, as this is the information at hand when setting the machine, see also Sec.~\ref{sec:appl1} and \cite{AABCR-2024,AFM2025}.}}, the pattern dataset $\boldsymbol \xi = \{\boldsymbol \xi^{\mu} \}_{\mu=1,...,K}$, that is quenched and drawn from a Rademacher distribution such that
\begin{equation} \label{eq:Rade}
\mathbb{P}(\xi_i^{\mu}) = \frac{1}{2} (\delta_{\xi_i^{\mu},+1} + \delta_{\xi_i^{\mu},-1});
\end{equation}
the control parameters $\lambda$ and $H$ that tune, respectively, the inter-layer interaction strength and the intensity of the external field. Also notice that, moving from \eqref{eq:hamiltonian} to \eqref{eq:HamHam}, we dropped the dependence on $\boldsymbol g$ as its specific structure \eqref{eq:g_struct} is intrinsically encoded by having split the pairwise interactions into the first and the third contributions on the right-hand-side of \eqref{eq:HamHam}.

To see the interplay between the contributions making up the cost function \eqref{eq:HamHam} (we recall that the first two contributions correspond to the sum of $L$ Hopfield models, while the third contribution introduces a coupling among them), let us set $H=0$ and notice that, in order to minimize the first contribution, the neurons in each layer tend to align with an arbitrary pattern, say $\boldsymbol \sigma^{a}=\boldsymbol \xi^{\mu}$, and, since patterns are (approximately\footnote{This follows from the choice \eqref{eq:Rade}, from which $ \red{N^{-1}}\boldsymbol \xi^{\mu} \cdot \boldsymbol \xi^{\nu} \approx \delta_{\mu,\nu}$, with negligible corrections in the limit $N \to \infty$.}) orthogonal, it follows that $m^a_{\mu}=1$ and $m^{a}_{\nu}=0$ for $\nu \neq \mu$; in order to minimize the third contribution, the pattern retrieved by different layers must be the same, apart from the sign\footnote{This intrinsic blemish will be fixed in Sec.~\ref{sec:NewModel} by adopting higher-order inter-later interactions, in such a way that the third contribution will as well favor the disentangled state.}: assuming $L$ even, $L/2$ layers are aligned with $\boldsymbol \xi^{\mu}$ and the other $L/2$ layers are aligned with $- \boldsymbol \xi^{\mu}$ in such a way that $\sum_{a=1}^{L}\sum_{b=1, b\neq a}^{L} m^a_{\mu} m^b_{\mu} = -L/2$ (when $L$ is odd, the unbalance makes the sum equal to $(L-1)/2$). Notice that the case where $\boldsymbol \sigma^a = \boldsymbol \xi^{\mu}$ and $\boldsymbol \sigma^{b}=\boldsymbol \xi^{\nu}$, with $\nu \neq \mu$ if $b \neq a$, minimizes the first contribution but is only a local minimum for the third contribution, which would approximately\footnote{Again, this follows from the choice \eqref{eq:Rade}.} equal zero.

\section{\red{Stationary state description}} \label{sec:model2}
\red{Before specifying the task that we aim to address with the model introduced above, we carry out a statistical mechanics investigation of the network's computational capabilities in order to get a description of its expected macroscopic behavior, once a stationary state is reached (at a given temperature $T=\beta^{-1}$).}
%
%The statistical-mechanics investigation of the model is 
This analysis is detailed in the App.~\ref{Appendix-Guerra} by exploiting interpolating techniques (see e.g., \cite{guerra2001sum,Agliari-NN2020}), while here we simply report the explicit expression of the quenched free energy $\mathcal{A}^{RS}$ found in the thermodynamic limit $N \to \infty$, under the replica-symmetry (RS) approximation and in the high-storage regime. Before presenting it, we anticipate that, beyond the Mattis magnetizations $\boldsymbol m^a$, for $a=1,...,L$, another set of macroscopic observables needs to be defined, that is, 
\begin{equation}
    q_{12}^{a} = \frac{1}{N} \sum_{i=1}^N {\sigma}_i^{a,(1)} \sigma_i^{a,(2)}, 
\end{equation}
which represents the overlap between the neural configurations of two replicas ${\boldsymbol \sigma}^{a,(1)}$ ${\boldsymbol \sigma}^{a,(2)}$, where the superscripts $(1)$ and $(2)$ denote the replica index. 
The above-mentioned RS approximation implies that, in the thermodynamic limit, the distribution of these macroscopic observables concentrates around their expectation values denoted as, respectively, $\bar m^a_\mu$ and $\bar q_{12}^a$ for $\mu=1,...,K$ and $a=1,...,L$.

Thus, we have
\begin{equation} \label{eq:A_RS}
\begin{array}{lll}
     \mathcal{A}^{RS}(\beta, \lambda, H, \boldsymbol h)&=& L\left(\log 2+\dfrac{\beta\gamma}{2}\right)+\SOMMA{a=1}{L}\mathbb{E}_{\bm\xi, x}\log\Bigg\{\cosh\Bigg[\SOMMA{\mu=1}{L}\beta\left(\SOMMA{b=1}{L}g_{ab}\bar{m}_\mu^{b}\right)\xi^\mu+\beta H \, h^{a}+x \sqrt{\beta \gamma \bar{p}_{12}^{a}}\Bigg]\Bigg\}
     \\\\
     && -\dfrac{\gamma}{2}\log(\mathrm{det}\mathcal{G})+\dfrac{\beta\gamma}{2}\SOMMA{a,b=1}{L} \sqrt{\bar{q}_{12}^{a}} (\mathcal{G}^{-1}\,)_{ab}\, \sqrt{\bar{q}_{12}^{b}}
     \\\\
     &&
     -\dfrac{\beta}{2}\SOMMA{a,b=1}{L}\SOMMA{\mu=1}{L}\bar{m}_\mu^{a}g_{ab}\bar{m}_\mu^{b}-\dfrac{\beta\gamma}{2}\SOMMA{a=1}{L}\bar{p}_{12}^{a}\Big(1-\bar{q}_{12}^{a}\Big)
\end{array}
\end{equation}
where $\gamma=\lim_{N\to\infty} K/N$ defines the network {\em storage},
\begin{equation}
         \bar{p}_{12}^{a}= -\SOMMA{\substack{b=1\\ b \neq a}}{L} \sqrt{\dfrac{\bar{q}_{12}^{b}}{\bar{q}_{12}^{a}}} \;(\mathcal{G}^{-1}\,)_{ab}\, -\SOMMA{c,b=1}{L} \sqrt{\bar{q}_{12}^{c}\bar{q}_{12}^{b}} \;\left[\partial_{\bar{q}_{12}^{a}}(\mathcal{G}^{-1}\,)_{cb}\, \right],
\end{equation}
%and
%$h^{(a)}(t))$ is the external field acting on the layers $a$ at time $t$ and
$\mathbb E_{\boldsymbol \xi, x}$ represents the quenched average over the realization of patterns and over the auxiliary standard-normal variable $x \sim \mathcal N(0,1)$, and 
\begin{equation}
   \mathcal{G}_{ab} =  \left(g^{-1}\right)_{ab}-\beta(1-\bar{q}_{12}^{a})\delta_{ab}
\end{equation}
which is well-defined since $\mathbf{g}$ is positive defined. %which occurs if $0\leq \lambda\leq 1/(L-1)$
\\
The expectation value of the order parameters appearing in the expression \eqref{eq:A_RS} can be obtained by extremizing $\mathcal{A}^{RS}(\beta, \lambda, H, \boldsymbol h)$ with respect to these parameters, resulting in the following self-consistency equations
\begin{equation} \label{eq:self}
\begin{array}{lll}
     \bar{m}_{\mu}^{a}&=&  \mathbb{E}_{\bm\xi, x}\left\{\tanh\left[\SOMMA{\nu=1}{L}\beta\left(\SOMMA{b=1}{L}g_{ab}\bar{m}_\nu^{b}\right)\xi^\nu+\beta H \,h^{a}+x \sqrt{\beta \gamma \bar{p}_{12}^{a}}\right]\xi^\mu\right\},
\\\\
     \bar{q}_{12}^{a}&=&  \mathbb{E}_{\bm\xi, x}\left\{\tanh^2\Bigg[\SOMMA{\nu=1}{L}\beta\left(\SOMMA{b=1}{L}g_{ab}\bar{m}_\nu^{b}\right)\xi^\nu+\beta H \, h^{a}+x \sqrt{\beta \gamma \bar{p}_{12}^{a}}\Bigg]\right\}.
\end{array}
\end{equation}
Although these expressions look fairly standard, when the expectation $ \mathbb{E}_{\bm\xi, x}$ is implemented, they become rather cumbersome, see for instance App.~\ref{Appendix-Guerra} and App.~\ref{subsec:explicit_L3}. For this reason, their numerical solution will be limited to the low-storage regime ($\gamma=0$), see Sec.~\ref{sec:ssr}.

\section{Disentangling spurious states}\label{sec:appl1}
The modular structure of an $L$-directional associative memory, can be leveraged to tackle several kinds of task beyond standard pattern retrieval. For instance, in \cite{AABCR-2024}, we considered pattern disentanglement in the case where patterns retrievable by different layers were independent. Dropping this independence condition makes the task more challenging and, in the current work, we deepen such a scenario. Specifically, we aim to exploit the model \eqref{eq:HamHam} for disentangling \emph{spurious states}, that is, we want to input information in the form of a mixtures of $L$ patterns (without loss of generality we consider the first $L$ patterns) as $\textrm{sign}(\boldsymbol \xi^1 + \boldsymbol \xi^2 + ... + \boldsymbol \xi^L)$ and to get as output the single components $\boldsymbol \xi^1, \boldsymbol \xi^2, ..., \boldsymbol \xi^L$, one per layer. In other words, we want the configurations $\boldsymbol \sigma^a = \boldsymbol \xi^{a}$ for $a=1,...,L$ (or any permutation that ensures that different layers retrieve all the different patterns in the input), referred to as $\boldsymbol \sigma^{(1,2,...,L)}$, to be stable and attracting the configuration $\boldsymbol \sigma^a = \textrm{sgn}(\boldsymbol \xi^1 + \boldsymbol \xi^2 + ... + \boldsymbol \xi^L)$ for $a=1,...,L$. Given this task, a natural choice for 
the field acting on each layer is 
\begin{equation} \label{eq:input}
h_i =  \mathrm{sign}\left(\sum_{\mu=1}^L \xi_i^{\mu}\right),~~ \textrm{for}~~ i=1,...,N. 
\end{equation}
Notice that this field is layer independent \red{hence the superscript $a$ has been dropped}.

%We emphasize that the presence of this field is necessary because, as explained in the previous section, when $H=0$, the system would be more likely to end up in staggered configurations like $(+\boldsymbol \xi^1, - \boldsymbol \xi^1,+\boldsymbol \xi^1, ...)$ or, possibly, to configurations where patterns not belonging to the initial mixture are retrieved.
%
%$h^{1}_i(t)=h^{(2)}_i(t)=h^{(3)}_i(t)=\mathrm{sign}\left[\xi_i^1+\xi_i^2+\xi_i^3\right]$ will be our clamped input.

The evolution towards the target configuration $\boldsymbol \sigma^{(1,2,...,L)}$ can be checked by different means. In particular, in Secs.~\ref{sec:HLN}-\ref{sec:LLNoisy}, we analytically investigate whether the latter corresponds to the stationary configuration resulting from the self-consistent equations \eqref{eq:self}, when the fields \eqref{eq:input} are applied. Next, in Secs.~\ref{sec:ssr}-\ref{sec:MC}, we numerically investigate whether, starting from the input configuration $\boldsymbol \sigma^a = \boldsymbol h$ for $a=1,...,L$, \red{referred to as $\boldsymbol \sigma^{(h)}$}, and applying the stochastic local-field-alignment (see, e.g., \cite{Amit}), the system eventually reaches the target configuration and this is stable. We recall that the stochastic local-field-alignment plays as neural dynamics for the network and reads 
\begin{eqnarray}
\label{eq:evolv}
%\sigma^{a}_i(t+1) &=& \textrm{sign} [\tanh(\beta\tilde{h}_i^{a}(t)) + x_i^{a}(t)] \\
\sigma^{a}_i(t+1) &=& \textrm{sign} [\tilde{h}_i^{a}(t) +  \beta^{-1} \zeta_i^{a}(t)] \\
      \tilde{h}_i^{a}(t) &=& \dfrac{1}{N}\SOMMA{b=1}{L}g_{ab}\SOMMA{\mu=1}{K}\SOMMA{j=1}{N}\xi_j^\mu \sigma_j^b(t) \xi_i^\mu+H h_i
\end{eqnarray}
where $t$ denotes the time step, $\zeta_i^{a}(t)$ is a stochastic contribution\footnote{We will set $\zeta = \textrm{atanh}(x)$ with $x$ a uniform random variable ranging in $[-1,+1]$; this choice ensures that the dynamics \eqref{eq:evolv} yields to a Boltzmann-Gibbs stationary state, such that this network can be seen as a \emph{generalized Boltzmann machine} \cite{Coolen}.} 
%$x_i^{a}(t)$ is a stochastic contribution\footnote{Specifically, $x$ is a uniform random variable ranging in $[-1,1]$; this choice ensures that the dynamics \eqref{eq:evolv} yields to a Boltzmann-Gibbs equilibrium.} 
and $\boldsymbol{\tilde h}^a \in \mathbb R^N$ is the local field acting on neurons in the $a$-th layer (stemming from the interactions with other neurons and from the external field).

\subsection{Stability analysis in the high-load, noiseless regime} \label{sec:HLN}
As mentioned in Sec.~\ref{sec:model}, the configuration $\boldsymbol \sigma^{(1,2,...,L)}$ where different layers retrieve different patters is only possible minimum (out of many) for the cost function \eqref{eq:HamHam}. Thus, before inspecting the ability of the model to disentangle spurious states, it is worth taking a look at some representative extremal configurations and at their stability in the noiseless scenario ($\beta \to \infty$). 
Keeping $L=3$, we focus on the following classes of neuronal configurations\footnote{We are referring to ``classes'' of neural configurations, because, beyond the degeneracy due to the permutation of the three patterns over the three layers, there is also a degeneracy due to the symmetry of the cost function \eqref{eq:HamHam} under spin flip of all the three layers.}: 
$$\boldsymbol \sigma^{(1,2,3)} = (\boldsymbol \xi^1, \boldsymbol \xi^2, \boldsymbol \xi^3)$$ 
$$\boldsymbol \sigma^{(1,1,1)} = (\boldsymbol \xi^1, \boldsymbol \xi^1, \boldsymbol \xi^1)$$
$$\boldsymbol \sigma^{(1,1,1')} = (\boldsymbol \xi^1, \boldsymbol \xi^1, -\boldsymbol \xi^1)$$
$$\boldsymbol \sigma^{(h)} = (\boldsymbol h, \boldsymbol h, \boldsymbol h),$$
where we recall that $\boldsymbol{h}$ is defined in \eqref{eq:input}.
\newline
For each of them we will estimate the energy, the consistency and the stability \red{for a fixed value of the network storage $\gamma \in \mathbb{R}^+$}.
Before proceeding, a couple of remarks are in order.
First, the previous neuronal states have been chosen because they are recognized to minimize at least one of the contributions making up the cost function \eqref{eq:HamHam} and, in fact, we checked that they are also solutions of the self-consistency equations. However, we recall that the last condition only ensures that these configurations are extremal for the free energy, but not necessarily minima, that is, stable points.
%: this prompts the study of the free-energy Hessian performed in Sec.~\ref{sec:LLNoisy}.
Second, here the consistency analysis is pursued by 
%
%Let us start with the inspection of the noiseless case, where, %recalling the Hamiltonian \eqref{eq:HamHam}, 
%
%
recalling the stochastic dynamics \eqref{eq:evolv}, \red{setting $\beta^{-1}=0$}, and checking whether the configurations remain unchanged, that is, recasting \eqref{eq:evolv} into an evolutionary rule for the Mattis magnetizations 
\begin{equation} \label{eq:magn_evolv}
    m^a_\mu(t+1)=\dfrac{1}{N}\SOMMA{i=1}{N}\xi_i^\mu \sigma_i^{a}(t+1)=\dfrac{1}{N}\SOMMA{i=1}{N}\xi_i^\mu \sigma_i^{a}(t)\mathrm{sign}\left[\tilde{h}_i^{a} \sigma_i^{a}(t)\right] ~~ \textrm{for}~~ a=1,...,L,
\end{equation}
we verify if they remain constant in time (e.g., moving from $t=0$ to $t=1$). The stability of these configurations is then examined computationally by checking whether these configurations are fixed-point attractors with a non-vanishing attraction basin.

%\begin{equation}
%\begin{array}{lll}
%      \tilde{h}_i^{(x)}(x,y,z) &=& \dfrac{1}{N}\SOMMA{\mu=1}{K}\SOMMA{j=1}{N}\xi_j^\mu x_j\xi_i^\mu+J h^{(1)}_i(t)-\dfrac{1}{N}\dfrac{\lambda}{2}\SOMMA{\mu=1}{K}\SOMMA{j=1}%{N}\xi_j^\mu\left(y_j+z_j\right)\xi_i^\mu
%\end{array}
%\end{equation}
%we have
%\begin{equation}
%    m_\mu^{(L+1)}=\dfrac{1}{N}\SOMMA{i=1}{N}\xi_i^\mu x_i^{(L+1)}=\dfrac{1}{N}\SOMMA{i=1}{N}\xi_i^\mu %x_i^{(L)}\mathrm{sign}\left[\tilde{h}_i^{(x)}(\bm x^{(L)},\bm y^{(L)},\bm z^{(L)})x_i^{(L)}\right]
%\end{equation}

The analytical estimates for \red{the one-step magnetization \eqref{eq:magn_evolv} and for the energy \eqref{eq:HamHam} related to the} four configurations above \red{follow from} straightforward but pretty lengthy \red{calculations that} are detailed in App. \ref{sec:consistency}, \red{while here we provide a summary of the results in Figure~\ref{fig:c_plot}.}

 \begin{figure}
     \centering
     \includegraphics[width=15cm]{PLOT/varie_conf_stability-4_v2.pdf}
     \caption{\red{We initialize the system in the configuration
     $\bm \sigma^{(1,2,3)}$ (first line), $\bm \sigma^{(1,1,1)}$ (second line), $\bm \sigma^{(1,1,1')}$ (third line), and $\bm \sigma^{(h)}$ (fourth line) and we evaluate analytically the related one-step magnetization \eqref{eq:magn_evolv}, thereby deriving the stability region (black line) in the $(H,\lambda)$ plane for that solution.} Specifically, these \red{lines} are obtained \red{by setting $\gamma=0.01$ and} by determining in which region of the plane the one-step magnetization (i.e., the error functions in, respectively, eqs.~\eqref{eq:erf_123}, \eqref{eq:erf_111}, \eqref{eq:erf_11a},~\eqref{eq:erf_11b}, \eqref{eq:erf_h}), exceed a certain threshold, which we set to $0.95$; for $\bm \sigma^{(h)}$ no boundaries are detected in the region under consideration.
     The shade in the color accounts for the energy associated to the related fixed point: the smaller the energy and the darker the color, \red{see also eqs.~\eqref{eq:HHH0}, \eqref{eq:HHH}, \eqref{eq:HHH11_1}, \eqref{eq:HHH1}}. Thus, for small $H$, although $\bm \sigma^{(1,2,3)}$ turns out to be stable, its energy is relatively close to zero. \red{These analytical predictions are validated against computational results in order to assess the configuration stabilities versus small perturbations. To this purpose, we} initialize the system in a configuration obtained 
     from $\bm \sigma^{(1,2,3)}$ (first line), from $\bm \sigma^{(1,1,1)}$ (second line), from  $\bm \sigma^{(1,1,1')}$ (third line), and from  $\bm \sigma^{(h)}$ (fourth line), by flipping randomly its entries: the flip is implemented by multiplying each neuron variable $\sigma_i^a$ by a random variable $\chi_i^a$ drawn from $P(\chi)= \frac{1+r}{2}\delta(\chi-1)+\frac{1-r}{2}\delta(\chi+1)$, where $r=1.0$ (left column), $r=0.8$ (middle column), and $r=0.5$ (right column), clearly, the larger $r$ and the closer the initial configuration to the reference. Then, we implement the dynamics \eqref{eq:evolv} with $T=0$, up to convergence to a fixed point. This is repeated for several choices of the parameters $H$ and $\lambda$ sampled uniformly in, respectively, $[0,2]$ and $[0,0.5]$ and for fixed $N=5000$ and $K=50$. Different final states are recorded and represented by different symbols and colors, as reported by the legend on the right: $\bm \sigma^{(1,2,3)}$ (green $\times$), $\bm \sigma^{(1,1,1)}$ (blue $+$), $\bm \sigma^{(1,1,1')}$ (magenta $\circ$), $\bm \sigma^{(h)}$ (red $\triangle$), or none of those considered in this section (gray $\bullet$). \red{The patterns presented in the figure are just for illustrative purposes as both analytical and numerical results are obtained for a Rademacher dataset; for an analysis involving structured data we refer to Sec.~\ref{sec:MC}.}
     %The stability range for the four examined configurations, predicted analytically by studying the Mattis magnetization evolution \eqref{eq:erf_123}, is represented by the black lines. Specifically, these are obtained by determining in which region of the ($\lambda$, $H$) plane the one-step magnetization (i.e., the error functions in, respectively, eqs.~\eqref{eq:erf_123},~\eqref{eq:erf_111},~\eqref{eq:erf_11a},~\eqref{eq:erf_11b},~\eqref{eq:erf_h}), exceed a certain threshold, which we set to $0.95$; for $\bm \sigma^{(h)}$ no boundaries are detected in the region under consideration.
     %The shade in the color accounts for the energy associated to the related fixed point: the smaller the energy and the darker the color. Thus, for small $H$, although $\bm \sigma^{(1,2,3)}$ turns out to be stable, its energy is relatively close to zero.
     }
     \label{fig:c_plot}
 \end{figure}
%%

%Here, the ranges of stability suggested by these analytical computations are presented in Figure~\ref{fig:c_plot} and corroborated by numerical tests. 
From this analysis it turns out that the configuration $\boldsymbol \sigma^{(1,2,3)}$ we are interested in is stable for relatively small values of $\lambda$ and of $H$, corresponding to the region highlighted by the green crosses in Figure~\ref{fig:c_plot} (first row). However, this state represents only a local minimum in the energy landscape and, if we initiate the dynamics from a different initial state, we may no longer converge to $\boldsymbol{\sigma}^{(1,2,3)}$, as shown in Figure~\ref{fig:c_plot} (second to fourth rows). Also, in this noiseless scenario, the configuration $\bm\sigma^{(h)}$ turns out to be stable for any choice of the parameters $\lambda$ and $H$, thus, some degree of noise is in order for this model to disentangle mixtures. This constitutes an analogy with the standard Hopfield modules, where odd mixtures like $\textrm{sign}(\boldsymbol \xi^1 + \dots +\boldsymbol \xi^{2n+1})$, with $n \in \mathbb N$, result to be stable at sufficiently low temperatures, thus the application of a certain degree of noise ($\beta^{-1}>0$) is a useful strategy to avoid these ``errors'' \cite{Amit}. Here the configuration $\boldsymbol \sigma^{(h)} = \textrm{sign}(\boldsymbol \xi^1 + \dots +\boldsymbol \xi^{L})$ is as well a fixed point and the application of some noise allows the system to escape its attractiveness and possibly move towards $\boldsymbol \sigma^{(1,\dots,L)}$. \red{The previous analysis was carried out for a fixed value of the network storage $\gamma = 0.01$, however, extensive simulations were performed for different values of relatively low values of $\gamma$, demonstrating the overall robustness of the network's dsentangling capability with respect to variations in network storage (see also App.~\ref{sec:NewModel}).}

\subsection{Stability analysis in the low-load, noisy, and zero-field regime}\label{sec:LLNoisy}

\begin{figure}[tb]
    \centering
    \includegraphics[width=0.45\linewidth]{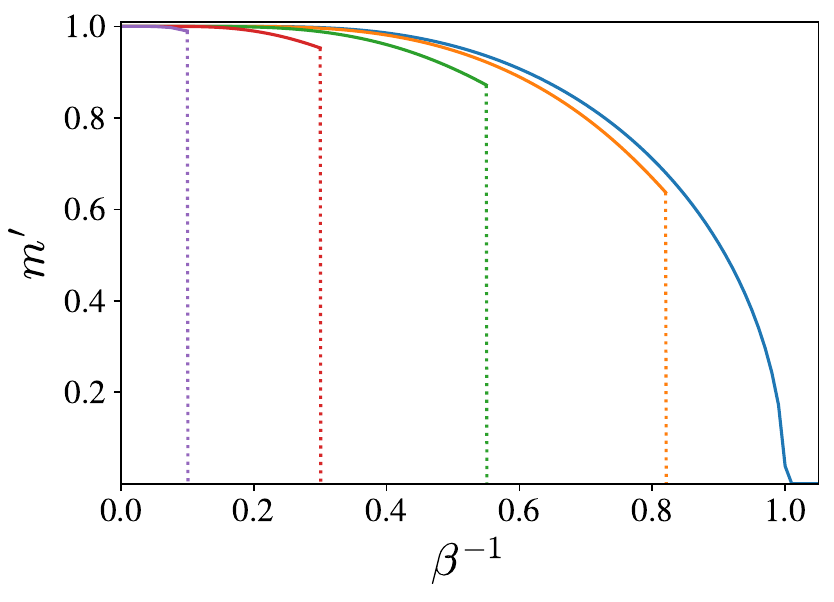}
    \includegraphics[width=0.45\linewidth]{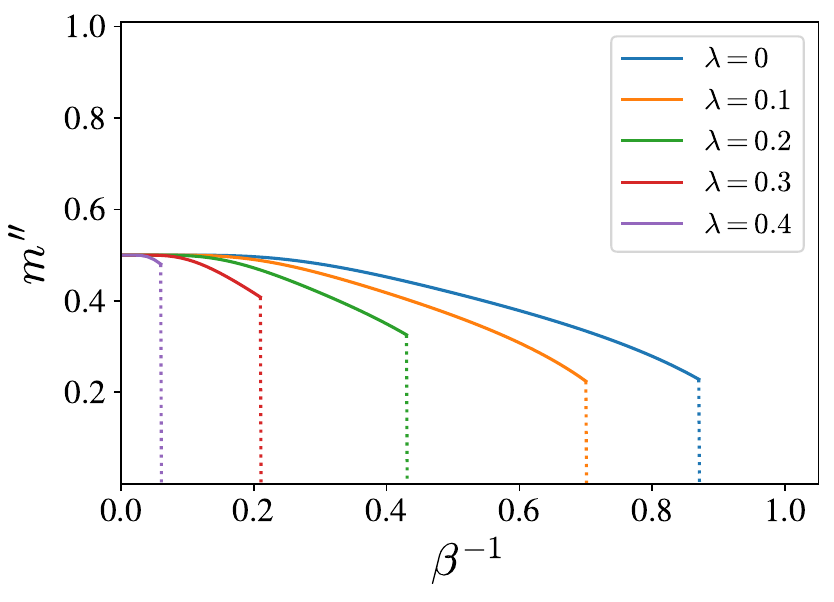}
    \caption{The solid lines represent the numerical solution of the self-consistency equations \eqref{eq:self} in the low-load regime and in the absence of external field, obtained by applying the fixed-point iteration method with initial point given by $\bar{\boldsymbol m}_{\{\mu \leq L\}}^{(1,2,3)}$ (left) and by $\bar{\boldsymbol m}_{\{\mu\leq L\}}^{(h)}$ (right), see eqs.~\eqref{eq:mm1}-\eqref{eq:mm2}. These numerical solutions preserve the structure of the initial datum, specifically, on the left, the solid lines show the behavior of $\bar{m}_1^1=\bar{m}_2^2=\bar{m}_3^3$ while $\bar{m}_{\mu}^a$ is vanishing for $\mu \neq a$; on the right, the the solid lines show the behavior of $\bar m_{\mu}^{a}$, that coincides for any $a \in [1,2,3]$ and $\mu \in [1,2,3]$. 
    %while for other values of $a$ and $\mu$ the resulting magnetization is vanishing. 
    The persistency in the structure of the solution is lost at a certain value of $\beta^{-1}$, highlighted by the vertical dotted lines: beyond these values, that depend on $\lambda$ (see the common legend on the right), solutions with a different structure appear, and these correspond, for instance, to the state $\boldsymbol \sigma^{(1,1,1')}$. 
    %and therefore to a matrix structure where $\bar{m}_1^a \neq 0$, while $\bar{m}_{\mu\neq 1}^a \neq 0$ for any $a \in [1,2,3]$.
    }
    \label{fig:enter-label}
\end{figure}

%The self consistent equations \eqref{eq:self} are mathematically derived by studying the saddle point equations $\nabla_{\{q,m\}} f_{RS} = 0$ of the RS free energy.
In this section, we set $\gamma =0$ and $H=0$, and \red{we focus on} two possible solutions of the saddle-point equations \eqref{eq:self}, that is, $\boldsymbol \sigma ^{(h)}$ and $\boldsymbol \sigma ^{(1,2,3)}$, corresponding to, respectively, the input and the target output of the disentanglement task under study. 
More precisely, we apply the fixed-point iteration technique to \eqref{eq:self}, by starting the procedure with the configurations $\boldsymbol \sigma^{(h)}$ and $\boldsymbol \sigma^{(1,2,3)}$. The related solutions are denoted by $\bar{\boldsymbol m}^{(h)}\in [-1, +1]^{K \times L}$ and $\bar{\boldsymbol m}^{(1,2,3)} \in [-1, +1]^{K \times L}$ and depicted in Figure~\ref{fig:enter-label}.

%We denote by $\bar{\boldsymbol m}^{(1,2,3)} \in [-1, +1]^{K \times L}$ and $\bar{\boldsymbol m}^{(h)}\in [-1, +1]^{K \times L}$ the matrices obtained by solving numerically the self-consistency equations \eqref{eq:self} in the low-load regime ($\gamma=0$) and in the absence of external field ($H=0$), and corresponding to configurations close to, respectively, $\boldsymbol \sigma^{(1,2,3)}$ and $\boldsymbol \sigma^{(h)}$, see Figure~\ref{fig:enter-label}.
%
We find that, as long as $\beta^{-1}$ is small enough, the following sub-matrices\footnote{The subscript $\{\mu \leq L \}$ highlights that we are focusing on the block with $\mu \leq L$ and the neglected entries are set equal to 0.}
%\begin{equation} \label{eq:mm1}
%\bar{\boldsymbol m}_{\{\mu \leq L\}}^{(1,2,3)}= 
%  \begin{bmatrix}
%    1 & 0 & 0\\
%    0 & 1 & 0\\
%    0 & 0 & 1\\
%  \end{bmatrix}
%\end{equation} 
%$\overline m_{\mu,a}= z \delta_{\mu a}$ (for $\mu\leq P$
%and 
%\begin{equation} \label{eq:mm2}
%\overline m_{\{\mu,a \leq L\}} = \overline %m_{\boldsymbol{\sigma}^{(\boldsymbol{h})}} = 
%\bar{\boldsymbol m}_{\{\mu\leq L\}}^{(h)}=
%  \begin{bmatrix}
%    0.5 & 0.5 & 0.5\\
%    0.5 & 0.5 & 0.5\\
%    0.5 & 0.5 & 0.5\\
%  \end{bmatrix},
%\end{equation} 
%
\begin{equation} \label{eq:mm1}
\bar{\boldsymbol m}_{\{\mu \leq L\}}^{(1,2,3)}= m'
  \begin{bmatrix}
    1 & 0 & 0\\
    0 & 1 & 0\\
    0 & 0 & 1\\
  \end{bmatrix}
\end{equation} 
%$\overline m_{\mu,a}= z \delta_{\mu a}$ (for $\mu\leq P$
and 
\begin{equation} \label{eq:mm2}
%\overline m_{\{\mu,a \leq L\}} = \overline m_{\boldsymbol{\sigma}^{(\boldsymbol{h})}} = 
\bar{\boldsymbol m}_{\{\mu\leq L\}}^{(h)}= m''
  \begin{bmatrix}
    1 & 1 & 1\\
    1 & 1 & 1\\
    1 & 1 & 1\\
  \end{bmatrix},
\end{equation} 
are fixed points for the equation \eqref{eq:self}, with the scalars $m'$ and $m''$ depending, in general, on $\beta$ and $\lambda$. As $\beta^{-1} \to 0$, $m'=1$ and $m''=0.5$, in such a way that $\bar{\boldsymbol m}_{\{\mu \leq L\}}^{(1,2,3)}$ and $\bar{\boldsymbol m}_{\{\mu\leq L\}}^{(h)}$ sharply correspond to \red{the magnetizations related to the configurations} $\boldsymbol \sigma^{(1,2,3)}$ and $\boldsymbol \sigma^{(h)}$, while, as $\beta^{-1}$ is increased, $m'$ and $m''$ progressively decrease, yet the matrix structure (scalar and constant) is fairly preserved; then, beyond a certain value of $\beta^{-1}$, we fail to find a solution with that kind of structure. This failure implies that extremal points nearby $\boldsymbol \sigma^{(1,2,3)}$ or $\boldsymbol \sigma^{(h)}$ (according to the magnetization matrix used for the initialization) no longer exist. \red{Remarkably, for a given $\lambda$ (e.g., $\lambda =0.2$) and spanning over larger and larger values of $\beta^{-1}$, this singularity occurs first for the input configuration  $\boldsymbol \sigma ^{(h)}$ ($\beta^{-1} \approx 0.45$) and then for the output configuration $\boldsymbol \sigma ^{(1,2,3)}$ ($\beta^{-1} \approx 0.55$).}

\red{As already recalled, the solutions of the self-consistency equations \eqref{eq:self} are not necessarily equilibrium states as we also need to check that these extremal states are minima of the free energy $f=-\beta \mathcal A$. We now proceed in this direction and we denote with}
\begin{align}
    D_{\mu\nu}^{ab} = \frac{\partial^2 f_{RS}}{\partial \overline m_\mu^a \partial \overline m_\nu^b}
\end{align}
\red{the entries of the Hessian matrix related to the free energy \eqref{eq:A_RS}, recalling that here $\gamma=0$.
Next, we determine the conditions on the control parameters $\beta$ and $\lambda$ under which the Hessian matrix is positive definite when evaluated at the magnetizations $\bar{\boldsymbol m}^{(1,2,3)}$ and $\bar{\boldsymbol m}^{(h)}$.
%Let us now focus on the stability of these solutions: 
%As we will show, it turns out that the related configurations $\boldsymbol{\sigma}^{(1,2,3)}$ and $\boldsymbol{\sigma}^{(h)}$ display different stability curves in the $(\beta,\lambda)$ plane and, in particular, there exists a non-empty region in the $(\beta, \lambda)$ plane, where only the diagonal solution $\boldsymbol{\sigma}^{(1,2,3)}$ is stable -- but, of course, there could be other ``spurious" states that can be stable in this region, making the disentanglement less efficient.
}

Starting from the second-order derivative
\begin{align}\label{eq:hessian}
    D_{\mu\nu}^{ab} = g_{ab} \delta_{\mu\nu} - \beta \sum_{c=1}^{L} g_{cb}g_{ca} \mathbb{E}_{\bm\xi}\left\{ \xi^\mu\xi^\nu \left[ 1-\tanh^2\lr{\beta \sum_{\rho=1}^{ L} \xi^\rho \sum_{d=1}^L g_{cd} \overline m_\rho^d} \right] \right\}
\end{align}
\red{and, following some straightforward algebraic manipulations (see App.~\ref{sec:spectrum}) we get the spectrum of the Hessian matrix. The stability of an extremal state depends on the sign of the smallest eigenvalue: if it is positive the solution is a minimum of the free energy $f_{RS}$ and therefore is said to be stable; otherwise, if negative, the solution is a saddle point or a maximum, and it is said to be unstable.}

\begin{figure}
     \centering
     \includegraphics[width=7.5cm]{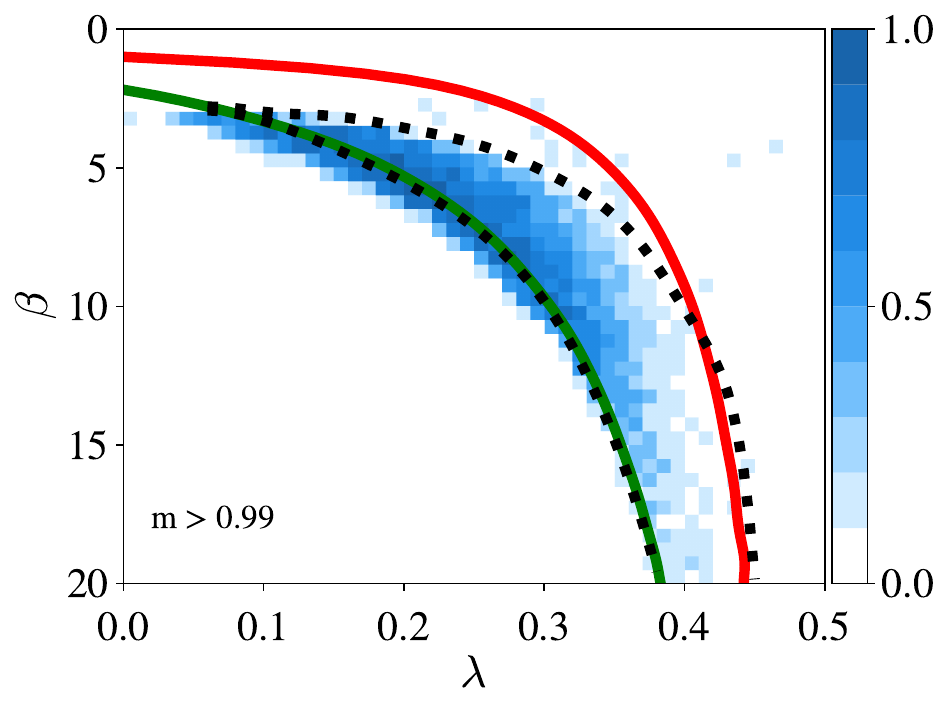}
     \includegraphics[width=7.5cm]{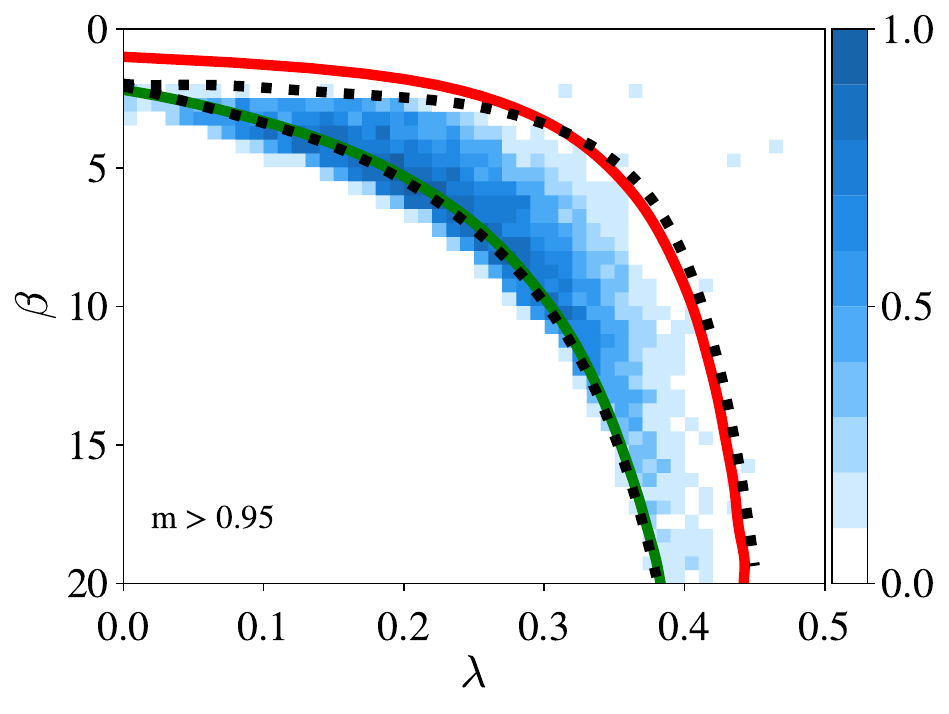}
     \caption{Both panels present the range in the parameter space $(\beta, \lambda, H=0)$ where the three-layer model is expected to work as pattern disentangler. Below the red line the target configuration $\boldsymbol \sigma^{(1,2,3)}$ is stable, while above the green line the spurious configuration $\boldsymbol \sigma^{(h)}$ is unstable. The two lines are found by studying the sign of the Hessian $D_{\mu \nu}^{aa}$, obtained for $N \to \infty$ and $\gamma=0$, as reported in Sec.~\ref{sec:LLNoisy} \red{and App.~\ref{sec:spectrum}}.  
     The dashed lines are found by solving the self-consistency equations \eqref{eq:self}, by the fixed-point iteration method, starting from $\boldsymbol \sigma^{(h)}$, as explained in Sec.~\ref{sec:ssr}. More precisely, in the region between the two dashed curves, the solution found in this way corresponds to $\boldsymbol \sigma^{(1,2,3)}$, therefore in that region we expect that the machine can successfully work. Notice that the region determined by this method is, consistently, within the region outlined by stability analysis and, since it is derived from the self-consistency equations holding under the RS assumption and in the thermodynamics limit, it is expected to be subject to the same conditions.
     As a final test, useful to check possible finite-size corrections, we run MC simulations with a network made of $N=5000$ neurons and $K=5$ patterns, by initializing the system in the configuration $\boldsymbol \sigma^{(h)}$, updating it according to \eqref{eq:evolv}, and keeping track of whether the stable state corresponds or, still, it is strongly correlated with, $\boldsymbol \sigma^{(1,2,3)}$: if the experimental magnitudes $m_1^1$, $m_2^2$, and $m_3^3$ (or suitable permutations) are \emph{simultaneously} larger than $0.99$ (left panel) or than 0.95 (right panel), the experiment is considered successful. Such trial is repeated $50$ times, for several choices of the parameters $\beta$ and $\lambda$, estimating the accuracy as the fraction of successful trails versus the number of trials (see the colormap). We remark that an overall very good agreement among the theoretical predictions and the numerical outcomes is obtained. }
     \label{fig:allyoucan}
   \end{figure}

The stability lines for the configurations $\boldsymbol \sigma^{(1,2,3)}$ and $\boldsymbol \sigma^{(h)}$ are reported in Figure~\ref{fig:allyoucan}. It is worth stressing that there exists a non-vanishing region, where $\boldsymbol \sigma^{(h)}$ is unstable while $\boldsymbol \sigma^{(1,2,3)}$ is stable and the existence of such a region is a strictly necessary condition for this model to work. In fact, by initializing the system in $\boldsymbol \sigma^{(h)}$, we first want to move away from that state and eventually reach $\boldsymbol \sigma^{(1,2,3)}$ \red{-- but, of course, there could be other ``spurious" states that can be stable in this region, making the disentanglement less efficient.} Consistently with the analysis led in Sec.~\ref{sec:HLN}, for this to occur the noise must be strictly positive. We also emphasize that the region determined here constitutes only an upper-bound as the instability and stability of, respectively, $\boldsymbol \sigma^{(h)}$ and $\boldsymbol \sigma^{(1,2,3)}$ do not directly imply that the former belongs to the attraction basin of the latter, that is, along its evolution, the system may bump into other stable states and remain nearby.

\subsection{\red{Checking disentanglement properties by numerical solutions of the saddle-point equations}}
\label{sec:ssr}
%\red{The results presented in Figure \ref{fig:enter-label} and stemming %from the numerical solution of the self-consistency equations %\eqref{eq:self} provides us information on the values of the parameters %$\beta$ and $\lambda$ where neuronal configurations exhibiting the %structure \ref{eq:mm1} and \ref{eq:mm2} are equilibrium states for the %system.} 
%Before proceeding, a procedural remark is in order. In fact, 
For classical retrieval tasks, checking that the retrieval configuration is a solution of the saddle-point equation with a finite attraction basin, namely checking that it is a (local) minimum for the free-energy,
is enough to state that the machine performs pattern retrieval. This can be inspected by solving the saddle-point equation via the fixed-point iteration method, starting from a configuration ``close'' to the retrieval one, \red{as previously done in Sec.~\ref{sec:LLNoisy}.} On the other hand, this kind of procedure is not sufficient for the current task, that is, checking that the configuration $\boldsymbol \sigma^{(1,...,L)}$ is a (local) minimum for the free-energy is only a necessary condition here. Indeed, we need to require a stronger condition, namely, that the input configuration $\boldsymbol \sigma^{(h)}$ is unstable and belongs to the attraction basin of $\boldsymbol \sigma^{(1,...,L)}$. A possible way to check this is by looking for the solution of the saddle-point equation when the configuration $\boldsymbol \sigma^{(h)}$ is chosen as the starting point of the iterative method. Then, if that configuration constitutes a free-energy minimum, the fixed-point method will return $\boldsymbol \sigma^* = \boldsymbol \sigma^{(h)}$, otherwise, we expect that it will return the closest minimum, where the system is likely to end up.

\begin{figure}
    \centering
    \includegraphics[width=0.8\linewidth]{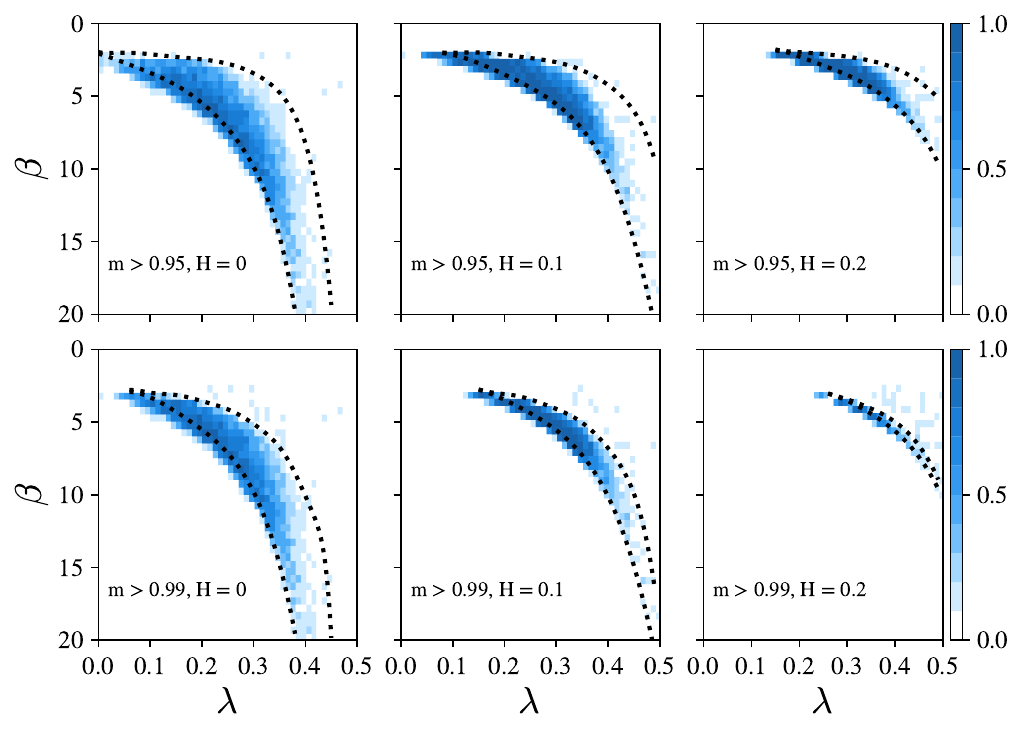}
    \caption{We estimate the region in plane $(\beta, \lambda)$, where the three-layer model is expected to successfully disentangle mixtures of three patterns by solving the self-consistent equations \eqref{eq:self} (dashed lines) and by running MC simulations (color map), in analogy to Figure~\ref{fig:allyoucan}; in both cases we considered several values of the external field $H=0.0$ (left column), $H=0.1$ (middle column), $H=0.2$ (right column), and two different thresholds on the magnetizations $m>0.95$ (upper row), $m>0.99$ (lower row). 
    For the first method, we set $\gamma =0$ and, as explained in \red{Sec.~\ref{sec:ssr}}, we found a region, bounded by the dashed lines, where the input configuration $\boldsymbol \sigma^{(h)}$ is attracted by the target output configuration $\boldsymbol \sigma^{(1,2,3)}$, thus within that region the system is expected to accomplish pattern disentanglement.
    For the second method, we set $N=5000$ and $K=5$, we initialize the system in the configuration $\boldsymbol \sigma^{(h)}$ and run the noisy dynamics \eqref{eq:evolv} up to convergence to a stationary state. Then, the magnetizations of the three layers versus the patterns $\boldsymbol \xi^1$, $\boldsymbol \xi^2$, $\boldsymbol \xi^3$, are evaluated and if each of the three patterns is retrieved with a quality at least equal to the given threshold (no matter which layer retrieves a certain pattern), the disentanglement achieved in that simulation is considered as successful. The accuracy is finally evaluated over the sample of $50$ trials and represented by the color map. }
    \label{fig:spurious-res}
\end{figure}

As mentioned in Sec.~\ref{sec:model}, the self-consistency equations \eqref{eq:self} are rather awkward and their numerical solution, following the protocol described above, is computationally demanding. Thus, we will focus on the low-load regime, where, under the simplifying assumption $\gamma=0$, more friendly expressions can be recovered, as detailed in App.~\ref{subsec:explicit_L3}.
The numerical solution of these self-consistency equations, setting $L=3$, is plotted in Figure~\ref{fig:allyoucan} and in Figure~\ref{fig:spurious-res} for different choices of $\beta$, $\lambda$ and $H$, and compared with the results obtained by studying the stability of $\boldsymbol \sigma^{(h)}$ and of $\boldsymbol \sigma^{(1,2,3)}$ (see the previous Sec.~\ref{sec:LLNoisy}) and with MC simulations (see the next Sec.~\ref{sec:MC}).
In particular, as $H$ gets larger, the successful region outlined by this method shrinks and moves toward larger values of $\lambda$ and smaller values of $\beta$, in fact, as $H$ gets larger the stability of the input configuration is reinforced, thus one needs a stronger inter-layer contribution and a higher degree of noise to destabilize it.

\subsection{\red{Checking disentanglement properties by Monte Carlo simulations}}
\label{sec:MC}

After the previous theoretically-driven analysis, we now tackle the problem computationally as this allows us to corroborate the theory, which is subjected to the RS and the thermodynamic limit assumptions. Moreover, the previous theoretically-driven analysis only provided an upper-bound for the region in the space $(\beta, \lambda, H)$ where we can expect the machine to work, without quantifying how well and how likely the machine can work. \red{Here, to answer this question, we run MC simulations, whose details, along with pseudo codes and a time consumption analysis are presented in App.~\ref{ssec:MC_details}.} 

\begin{figure}[t]
    \centering
    \includegraphics[width=15cm]{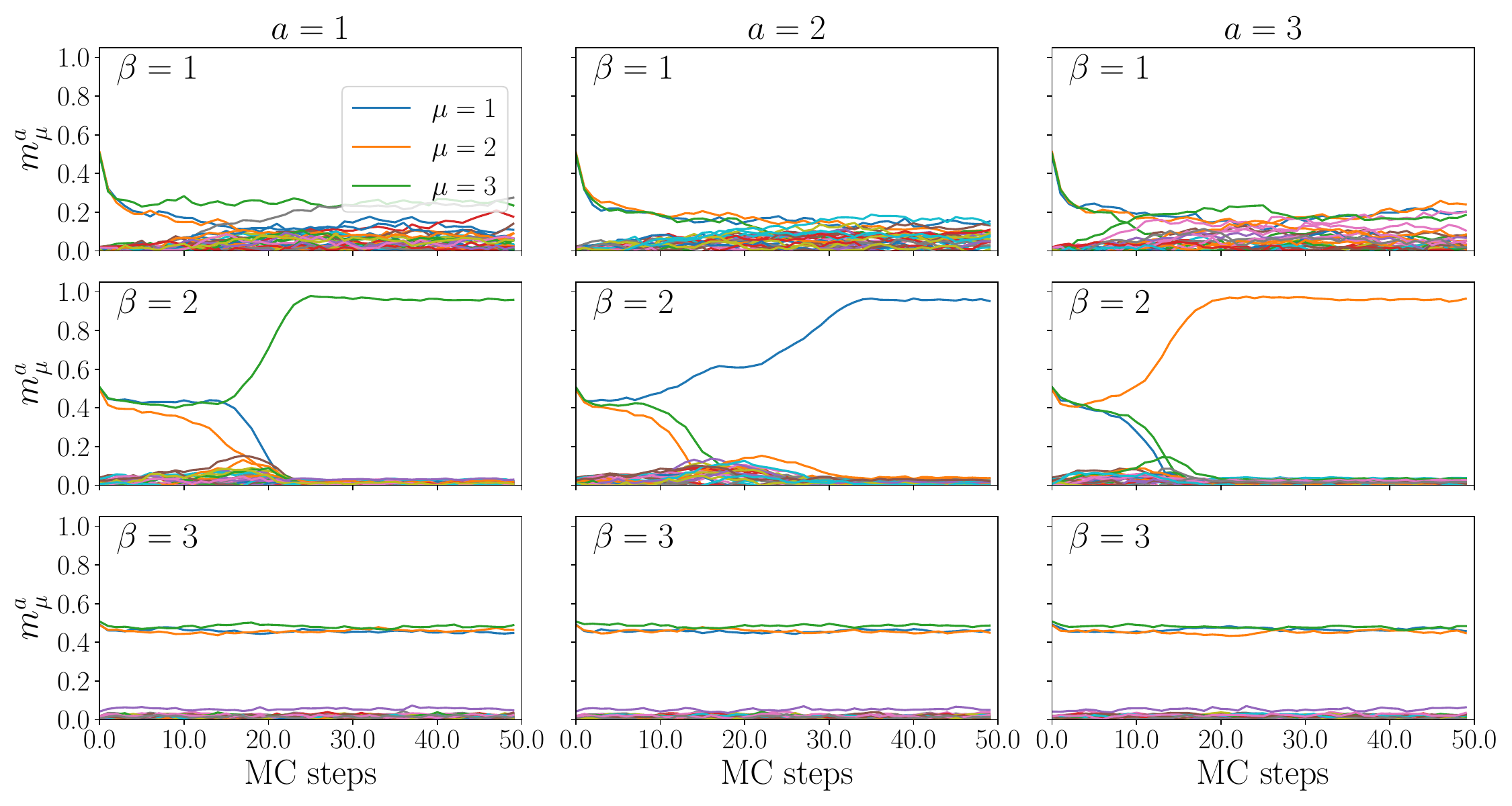}
    \caption{These plots show the evolution of the Mattis magnetizations $m_{\mu}^a$ for $\mu=1, ...,K$ (different labels correspond to different colors) and for $a=1, 2, 3$ (different layers correspond to different columns) versus the number of MC steps -- one MC step corresponds to $N$ random extractions of the index $i \in \{1, ..., N\}$ that identifies the neuron to be updated according to the rule \eqref{eq:evolv}\red{, see also App.~\ref{ssec:MC_details}.} More precisely, here we set $N=5000$, $K=50$ $H=0.2$ and $\lambda=0.2$, while different values of $\beta$ are chosen: $\beta=1$ (upper row), $\beta=2$ (middle row), $\beta=3$ (lower row); in agreement with the findings presented in Figure~\ref{fig:allyoucan}, the emerging behavior is, respectively, ergodic, disentangled, and stuck in the spurious state.}
    \label{fig:GTAM}
\end{figure}

\red{In our experiments} we initialize the system in the spurious state 
%$\boldsymbol \sigma = \textrm{sign}(\boldsymbol \xi^1 + \boldsymbol \xi^2 + \boldsymbol \xi^3)$, 
$\boldsymbol \sigma^{(h)}$, 
we let it evolve according to \eqref{eq:evolv} and, once a stable state is reached, we check whether this is retrieving the single components, that is, if it corresponds to $\boldsymbol \sigma^{(1,2,3)}$ (or any suitable permutation): \red{, see Figure~\ref{fig:GTAM} where  we inspect the time evolution of the magnetizations $m_1^1, m_2^2$, and $m_3^3$.} We repeat the experiment several times, \red{spanning over the parameters $\beta, \lambda, H$ and} counting the number of successful experiments, where ``successful'' means that the magnitudes of the observed magnetizations $m_1^1, m_2^2, m_3^3$ are larger than a certain threshold. Finally, the accuracy is evaluated as the fraction between the number of successful experiments and the overall number of experiments, and plotted in Figure~\ref{fig:allyoucan} and in Figure~\ref{fig:spurious-res}. Remarkably, there exists a region, inside the upper-bound determined analytically, where the accuracy is unitary or very close to one, and the existence of such a region guarantees that the machine can disentangle the inputted spurious state. Of course, this region gets wider as the threshold for success is lowered.

\red{The robustness of these results and their scalability versus $L$ is discussed in App.~\ref{sec:L5}, where the analysis for the case $L=5$ are reported.}

\begin{figure}[tb]
    \centering
    \includegraphics[width=14cm]{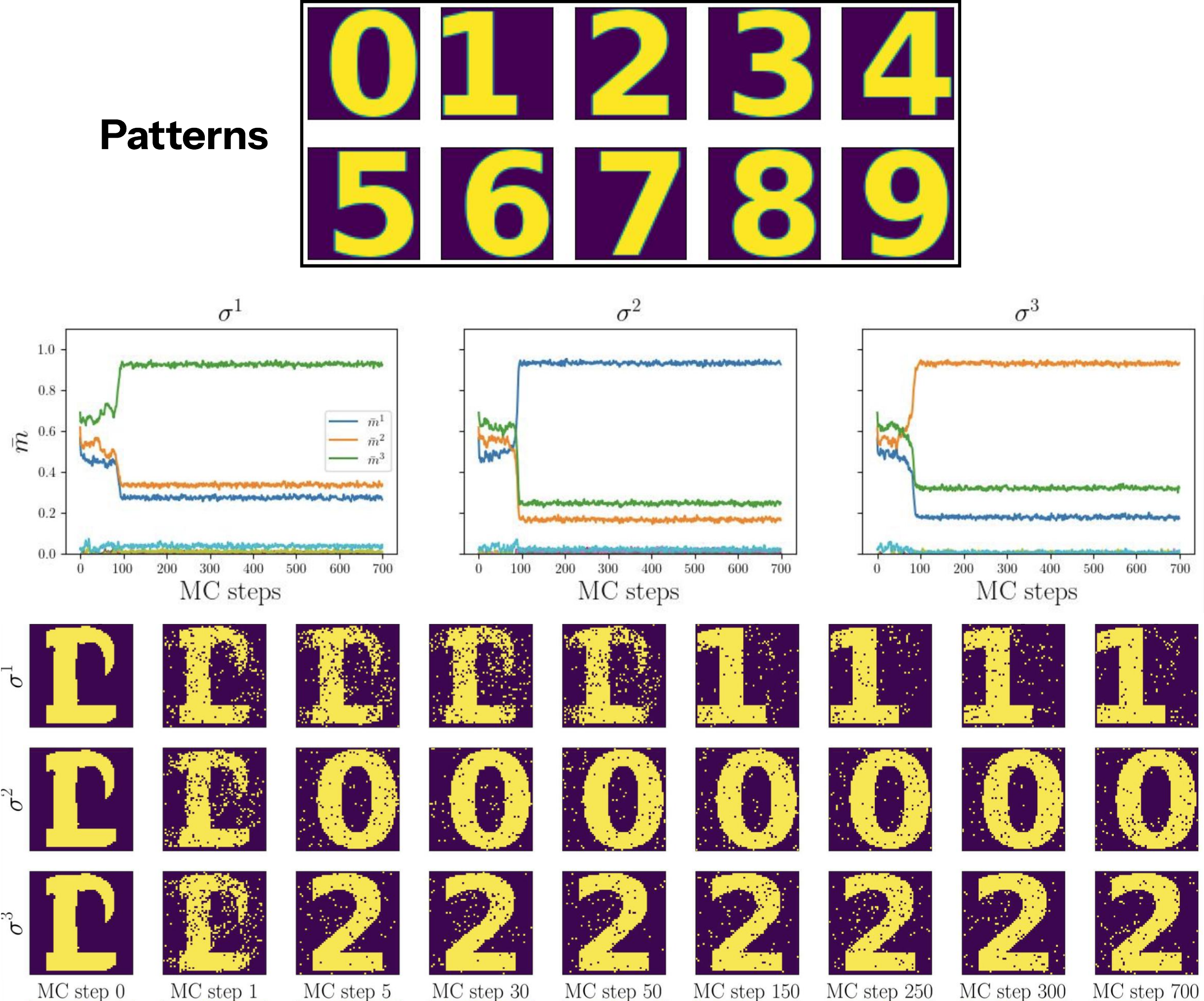}
    \caption{\red{We consider a dataset consisting of 10 digits, each represented by $56 \times 56$ pixels, see the upper part of the figure. This dataset is Hebbian-stored in a three-layer network governed by the cost function \eqref{eq:HamHam} with parameters $N= 3136$, $K=10$, $H=0.1$, and $\lambda =0.19$. A mixture of digits $0, 1, 2$ is then prepared and presented as input to each layer of the network, which is subsequently updated through MC simulations with $\beta =1.9$. The lower part of the figure illustrates the evolution of the neuronal configurations $\boldsymbol{\sigma}^1$, $\boldsymbol{\sigma}^2$, and $\boldsymbol{\sigma}^3$. Correspondingly, the middle part displays the evolution of the associated Mattis magnetizations $\boldsymbol{m}^1$, $\boldsymbol{m}^2$, and $\boldsymbol{m}^3$. Different colors are used to distinguish the magnetizations related to the different patterns comprising the dataset, with emphasis on the digits included in the input mixture, as highlighted in the legend.}}
    \label{fig:digiti}
\end{figure}

\red{We close this section by noting that, as we move away from the Rademacher dataset toward more structured patterns, the network's performance is expected to deteriorate. This is because Hebbian couplings are known to be particularly effective when the stored memories are (approximately) orthogonal. It is therefore worth considering a benchmark dataset to verify whether the disentanglement capabilities of the model \eqref{eq:HamHam} are preserved. In Figure \ref{fig:digiti}, we provide numerical evidence supporting this.
However, we emphasize that in this case the performance is more sensitive to parameter tuning, and the successful region in the $(\beta, \lambda)$ plane is expected to be smaller than in the Rademacher case\footnote{Also, in general, the optimal parameter setting might depend on the mixed patterns.}. This observation further underscores the importance of having a solid theoretical foundation.
We also note, consistently with the analytical results presented earlier (see, e.g., Sec.~\ref{sec:HLN}), that a certain degree of noise (i.e., $\beta^{-1} > 0$) is still required, implying that the retrieval of the deconvolved patterns exhibits some imperfections. To enable effective performance even in the noiseless regime, certain adjustments can be made — specifically, by introducing higher-order inter-layer interactions, as discussed in App.\ref{sec:NewModel}.}

\section{Conclusions} \label{sec:concl}
Triggered by the 2024 Nobel prize in Physics given to John Hopfield and Geoffrey Hinton for their pivotal contribution to the development of neural networks and learning machines, in this paper we verified Anderson's principle \cite{AndersonScience} on neural networks, by using as elements to be combined exactly Hopfield's neural networks \cite{Hopfield}. We therefore considered an assembly of $L$ Hopfield models, referred to as layers, each associated to the same dataset and coupled together. In this way, neurons are subject to intra-layer and inter-layer interactions that are both taken of Hebbian nature, however, while the former is ``imitative'' the latter is ``repulsive''.  
We showed that this kind of system exhibits capabilities that go beyond the classical pattern retrieval and which are not addressable by a single Hopfield model or even by an $L$-layer hetero-associative model displaying an analogous architecture \cite{Kosko,AABCR-2024}.
In fact, our model is able to disentangle mixtures of signals: if inputted with a composite information, it returns as output the single constituting signals.  

%
%that The structure of this system is analogous to that of hetero-associative associative memory
%
%introduced a $L$-layered associative neural network, where the Hebbian nature of neuronal interactions is retained while the underlying structure is designed in order to make the network able to perform tasks beyond the classical pattern retrieval \cite{Hopfield-1983} and/or association \cite{Kosko}. More precisely, our model is aimed at disentangling mixtures of signals: we want to input it with a composite information and to receive in output the single constituting signals.  
In particular, here, given a dataset of binary vectors $\boldsymbol \xi = \{\boldsymbol \xi^{\mu} \}_{\mu=1, ...,K} \in \{-1, +1\}^{N \times K}$, the input is given by mixtures like $\boldsymbol \sigma^{(h)} = \textrm{sign}(\boldsymbol \xi^1 + \boldsymbol \xi^2 + ... + \boldsymbol \xi^L)$ -- and this is interpreted as the initial neuronal configuration for each layer, that is, $\boldsymbol \sigma^a = \boldsymbol \sigma^{(h)}$ for $a=1, ..., L$ -- while the desired output is given by $\boldsymbol \sigma^{(1, 2, ...,L)}: \boldsymbol \sigma^{\ell} = \bm \xi^{\ell}$ for $\ell=1, ..., L$ (without loss of generality) -- and this is interpreted as the target state reached by the system. 

%To this goal, our layered associative neural network is equipped with positive intra-layer interactions and negative inter-layer interactions; an external field pointing in the direction of the input is also introduced. Thus, unlike previous layered models such as Kosko's BAM, here different layers display an anti-imitative coupling which is responsible for a more rugged energy landscape. In fact, before addressing the model's disentangling capabilities, we carried one 

We started our investigation with some preliminary analysis meant to secure the existence of a region in the space of control parameters where the configuration $\boldsymbol \sigma^{(h)}$ is unstable (as we do not want to remain stuck there), while the target configuration $\boldsymbol \sigma^{(1, 2, ...,L)}$ is stable. In fact, this is the case for intermediate values of the inter-layer coupling strength, not too large external fields and non-zero noise affecting the neuronal dynamics.   

Next, we solved for the free-energy of this model at the RS level of description and obtained a set of self-consistency equations for its order parameters. Given the non-classical task under study, the numerical solution of these equations also implies some adjustments: instead of checking that a certain configuration (typically, the retrieval configuration) is solution, we check that, inserting $\boldsymbol \sigma^{(h)}$ as candidate solution, the fixed-point interaction method converges to $\boldsymbol \sigma^{(1, 2, ...,L)}$. The results obtained in this way are perfectly consistent with the above-mentioned stability analysis.

Finally, we run MC simulations and corroborate the theoretically-driven results. Specifically, we are able to predict a proper setting for the control parameters of the model where the system is certainly able to perform the assigned task and a looser region where the system is very likely to perform the assigned task.

We emphasize that the kind of interactions implemented in this network yields a plethora of minima which can impair the disentanglement of the neuronal configuration $\boldsymbol \sigma^{(h)}$ into $\boldsymbol \sigma^{(1, 2, ...,L)}$. A way to see this is by considering an equivalent model obtained by applying a Hubbard-Stratonovich transformation to the model's partition function (see App.~\ref{Appendix-Guerra}) and notice that the interaction among the dummy variables $z$'s is characterized by a high degree of frustration, especially compared with other layered associative-memory models, see e.g., \cite{AABCR-2024}.
Many possible adjustments can be implemented to improve the performance of this model, for instance one can revise the Hebbian kernel to obtain a projection kernel \cite{Kohonen-1972,FAB-NN2019} that reduces the detrimental effects due to interference among the stored patterns, or allow for higher-order interactions \cite{FAB-NN2019, Krotov2018,Pedreschi1,Pedreschi2} which make the desired minima more stable. 
%The last strategy is implemented in the last part of the paper, specifically, the pair-wise \textit{heterogeneous} terms, namely those involving neurons from different layers, are replaced by fourth-order terms. Remarkably, this revision makes the model invariant under the spin-flip of a single layer, yields to more attractive disentangled states and therefore to a better performing model as far as the disentanglement task is concerned.
%%
%
\red{In fact, the architecture studied here opens several avenues. For instance, it would be natural to investigate the learning capabilities of Restricted Boltzmann Machines, which are equivalent to these modular networks (as highlighted in \cite{AABCR-2024}), or possibly to extend the framework to spiking neural networks (see, e.g., \cite{Shen_Xu_Liu_Wang_Pan_Tang_2023,Shen_Ni_Xu_Tang_2024,CinesiIngordi,xu2023enhancing}).}

\section*{Acknowledgments}
The authors are grateful to Alberto Fachechi and Paulo Duarte Mourão for useful discussions.\\ 
E.A. acknowledges financial support from PNRR MUR Project PE0000013-FAIR
and from Sapienza University of Rome (RM12117A8590B3FA, RM12218169691087).\\ 
A.B and E.A are members of GNFM-INdAM which is acknowledged.\\
A.B. and M.S.C. acknowledge PRIN 2022 Grant Statistical Mechanics of Learning Machines: from algorithmic and information theoretical limits to new biologically inspired paradigms n. 20229T9EAT funded by European Union—Next Generation EU. \\
The research has received financial support from the ‘National Centre for
HPC, Big Data and Quantum Computing—HPC’, Projects CN-00000013, CUP B83C22002940006, NRP Mission 4 Component 2 Investment 1.5, Funded by the European Union—NextGenerationEU.

\setcounter{equation}{0}
\renewcommand\theequation{A.\arabic{equation}}

\appendix

\section{RS solution by interpolation technique}\label{Appendix-Guerra}

Resuming the cost function \eqref{eq:HamHam}
\begin{equation}
    \mathcal{H}(\boldsymbol \sigma ; H,\boldsymbol g, \boldsymbol \xi, \boldsymbol h )= -\dfrac{N}{2} \SOMMA{\mu=1}{K}\SOMMA{a,b = 1}{L} m_\mu^ag_{ab}m_\mu^b - H \SOMMA{i=1}{N}\SOMMA{a=1}{L} h_i^a(t)\sigma_i^a
\end{equation}
where $g_{ab}= \delta_{ab}-\lambda(1-\delta_{ab})$, and under the condition $\lambda>\dfrac{1}{(L-1)}$, the partition function of the model reads as
\begin{equation}
\begin{array}{lll}
     \mathcal{Z}_N(\beta,H,\boldsymbol g, \boldsymbol \xi, \boldsymbol h )&=& \SOMMA{\{\bm\sigma^{1}\}}{}\cdots\SOMMA{\{\bm\sigma^{L}\}}{}\exp\Bigg[\dfrac{\beta N}{2} \SOMMA{\mu=1}{K}\SOMMA{a,b = 1}{L} m_\mu^ag_{ab}m_\mu^b +\beta H \SOMMA{i=1}{N}\SOMMA{a=1}{L} h_i^a(t)\sigma_i^a\Bigg].
\end{array}
\end{equation}
\red{For completeness, we also recall the full list of observables:
}
\begin{equation}
    \begin{array}{lll}
         \bar{m}_\mu^a =\mathbb{E} \dfrac{1}{N}\SOMMA{i=1}{N}\xi_i^\mu\omega(\sigma_i^a)&&\mathrm{with}\;\;\;a,\mu=1,\hdots, L
         \\\\
         q_{11}^{ab} = \dfrac{1}{N}\SOMMA{i=1}{N}\omega(\sigma_i^a\sigma_i^b,)&&q_{11}^{a} = 1,
         \\\\
         q_{12}^{ab} = \dfrac{1}{N}\SOMMA{i=1}{N}\omega(\sigma_i^a)\omega(\sigma_i^b),&&q_{12}^{a} = \dfrac{1}{N}\SOMMA{i=1}{N}\omega^2(\sigma_i^a),
         \\\\
         p_{11}^{ab} = \dfrac{1}{K-L}\SOMMA{\mu>L}{K}\omega(z_{\mu}^az_{\mu}^b),&&p_{11}^{a} = \dfrac{1}{K-L}\SOMMA{\mu>L}{K}\omega((z_{\mu}^a)^2),
         \\\\
         p_{12}^{ab} = \dfrac{1}{K-L}\SOMMA{\mu>L}{K}\omega(z_{\mu}^a)\omega(z_{\mu}^b)&&p_{12}^{a} = \dfrac{1}{K-L}\SOMMA{\mu>L}{K}\omega^2(z_{\mu}^a).
    \end{array}
\end{equation}
In the retrieval regime we ask the various layers to retrieve, exhaustively, the $L$ patterns making up the input mixture, that is, without loss of generality, we ask that $\boldsymbol \sigma^{\ell} = \xi^{\ell}$, for $\ell =1, ...,L$. Under these assumptions we are able to split the signal ($\mu\leq L$) from the noise terms ($\mu> L$) in the partition function:
\begin{equation}
\begin{array}{lll}
     \mathcal{Z}_N(\beta,H,\boldsymbol g, \boldsymbol \xi, \boldsymbol h)&=& \SOMMA{\{\bm\sigma^{1}\}}{}\cdots \SOMMA{\{\bm\sigma^{L}\}}{}\exp\Bigg[\dfrac{\beta N}{2} \SOMMA{\mu=1}{L}\SOMMA{a,b = 1}{L} m_\mu^ag_{ab}m_\mu^b+\dfrac{\beta N}{2} \SOMMA{\mu>L}{K}\SOMMA{a,b = 1}{L} m_\mu^ag_{ab}m_\mu^b +\beta H \SOMMA{i=1}{N}\SOMMA{a=1}{L} h_i^a(t)\sigma_i^a\Bigg]
\end{array}
\end{equation}
The noise term can be rewritten exploiting the $(K\times L)-$dimensional multivariate Gaussian transform, namely:
\begin{equation}
\begin{array}{lll}
     \mathcal{Z}_N(\beta,H,\boldsymbol g, \boldsymbol \xi, \boldsymbol h)&=& \SOMMA{\{\bm\sigma^{1}\}}{}\cdots \SOMMA{\{\bm\sigma^{L}\}}{}\displaystyle\int\mathcal{D}(z)\exp\Bigg[\dfrac{\beta N}{2} \SOMMA{\mu=1}{L}\SOMMA{a,b = 1}{L} m_\mu^ag_{ab}m_\mu^b +\beta H \SOMMA{i=1}{N}\SOMMA{a=1}{L} h_i^a(t)\sigma_i^a
     \\\\
     && +\sqrt{\beta N} \SOMMA{\mu>L}{K}\SOMMA{a = 1}{L} m_\mu^az_{\mu}^a\Bigg]
\end{array}
\end{equation}
where $\mathcal{D}(z)$ is the Gaussian measure with covariance $\boldsymbol g^{-1}$.
%we have used
%\begin{equation}
%    \displaystyle\int\mathcal{D}(z_{\mu}^a)=\displaystyle\int\prod\limits_{\mu>L,a,b=1}^{K,L}\dfrac{dz_{\%mu,a}dz_{\mu}^b}{2\pi\sqrt{\mathrm{det}g}}\exp\Bigg[-\dfrac{1}%{2}\SOMMA{\mu>L,a,b}{} z_{\mu}^a(g^{-1})_{ab}\,z_{\mu}^b\Bigg]
%\end{equation}
We compute the self-averaging statistical pressure $\mathcal{A}(\beta,  H, \bm g, \boldsymbol h)$, defined as
\begin{align}
\label{eq:def_stat_press}
    \mathcal{A}(\beta, H, \bm g,\boldsymbol h) = \lim_{N\to\infty} \frac{1}{N} \mathbb E \ln \mathcal{Z}_N(\beta,H,\boldsymbol g, \boldsymbol \xi, \boldsymbol h),
\end{align}
with the quenched expectation taken over the patterns $\xi^\mu$, by using the Guerra's interpolation method.
\red{The basic idea to compute the free energy within this framework, is to introduce an interpolating parameter $t \in [0,1]$ and a corresponding interpolating free energy function $\mathcal{A}(\beta, H, \bm g,\boldsymbol h; t)$. This function is constructed so that at $t = 1$, it coincides with the free energy of the original model, i.e., $\mathcal{A}(\beta, H, \bm g,\boldsymbol h; t=1) = \mathcal{A}(\beta, H, \bm g,\boldsymbol h)$. At the other endpoint, when $t = 0$, $\mathcal{A}(\beta, H, \bm g,\boldsymbol h;t=0)$ corresponds to the free energy of a simplified one-body system, where each neuron is decoupled from the rest and instead interacts with a properly designed external field. This field is tailored to reproduce, at least in terms of low-order statistics, the internal field that would be generated by the actual network.
\\
In defining the one-body system (i.e., at $t=0$), we also include auxiliary constants and functions, which we retain the flexibility to choose later in a way that simplifies the analysis and, in the thermodynamic limit  ($N\to\infty$), ensures Replica Symmetry (RS).
Under the RS ansatz, we assume that the probability distributions  of the order parameters become Dirac deltas in the thermodynamic limit, hence the expectations of the order parameters collapse on these values in this asymptotic limit, that is, calling $x$ a generic order parameter, $\lim_{N \to \infty}\langle x(\bm \sigma)\rangle = \bar{x}$.
\\
Ultimately, the central mathematical tool used here is the Fundamental Theorem of Calculus, which serves as a natural link between the two boundary cases of the interpolating parameter. This leads us to the sum rule that follows:
}
\begin{align}
\label{eq:calc_theorem}
    \mathcal{A}^{\red{RS}}(\beta, H, \bm g,\boldsymbol h) = \mathcal{A}^{\red{RS}}(\beta, H, \bm g,\boldsymbol h; t=0) + \int_0^1 dt \:\frac{\mathcal{A}^{\red{RS}}(\beta, H, \bm g,\boldsymbol h;s)}{ds}\Bigg\vert_{s=t},
\end{align}
with $\frac{\mathcal{A}^{\red{RS}}(\beta, H,\bm g, \boldsymbol h;t)}{dt}=\lim\limits_{N\to\infty} \frac{1}{N} \mathbb E\omega_t(\frac{\mathcal{Z}_N(\beta, H, \bm g,\boldsymbol h;t)}{dt})\equiv \lim\limits_{N\to\infty} \frac{1}{N} \avg{\frac{\mathcal{Z}_N(\beta, H,\bm g, \boldsymbol h;t))}{dt}}_t$, where we defined the quenched expectation over the (interpolating) Boltzmann average $\omega_t$ as 
\begin{align}
    \mathbb E\omega_t(.)\equiv \avg{.}_t,
\end{align}
which is taken over the interpolating measure:
\begin{equation}
\begin{array}{lll}
     \mathcal{Z}_N(\beta, H,\bm g, \boldsymbol h;t)&=&\SOMMA{\{\bm\sigma^{1}\}}{}\cdots \SOMMA{\{\bm\sigma^{L}\}}{} \displaystyle\int\mathcal{D}(z)\exp\Bigg[t\beta N \SOMMA{\mu=1}{L}\SOMMA{a,b = 1}{L} m_\mu^ag_{ab}m_\mu^b +\beta H \SOMMA{i=1}{N}\SOMMA{a=1}{L} h_i^a(t)\sigma_i^a+\sqrt{t}\sqrt{\dfrac{\beta}{N}} \SOMMA{\mu>L, i=1}{K,N}\SOMMA{a = 1}{L} \xi_i^\mu \sigma_i^a z_{\mu}^a
     \\\\
     &&  + (1-t)N \SOMMA{a=1}{L}\SOMMA{\mu=1}{L}\psi^{(a)}m_\mu^a+\sqrt{1-t}\SOMMA{\mu>L}{K}\tilde Y_\mu \SOMMA{a=1}{L}B^{(a)}z_{\mu}^a+\sqrt{1-t}\SOMMA{i=1}{N}Y_i \SOMMA{a=1}{L}A^{(a)}\sigma_i^a
     \\\\
     &&+\dfrac{1-t}{2}\SOMMA{i=1}{N} \SOMMA{\substack{a,b=1\\ a \neq b}}{L}C^{(ab)}\sigma_i^a\sigma_i^b+\dfrac{1-t}{2}\SOMMA{\mu>L}{K} \SOMMA{\substack{a,b=1}}{L}\tilde{C}^{(ab)}z_{\mu}^az_{\mu}^b\Bigg].
\end{array}
\label{eq:interp_part_general}
\end{equation}

the $t-$ derivative of $\mathcal{A}^{\red{RS}}(\beta, H, \bm g,\boldsymbol h;t)$, after we have set the interpolating constants as 
\begin{equation}
    \begin{array}{lll}
         (A^{(a)})^2=\beta\gamma \bar{p}_{12}^{a};&& A^{(a)}A^{(b)}=\beta\gamma \bar{p}_{12}^{ab};%=\red{\beta\gamma\sqrt{\bar{p}_{12}^a\bar{p}_{12}^b}}
         \\\\
         (B^{(a)})^2=\beta\bar{q}_{12}^{a};&& B^{(a)}B^{(b)}=\beta \bar{q}_{12}^{ab};%=\red{\beta\sqrt{\bar{q}_{12}^a\bar{q}_{12}^b}}
         \\\\
         C^{(a)}=\beta(1-\bar{q}_{12}^{a});&& \tilde{C}^{(ab)}=\beta(\bar{q}_{11}^{ab}-\bar{q}_{12}^{ab});%=\red{\beta(\sqrt{\bar{q}_{11}^a\bar{q}_{11}^b}-\sqrt{\bar{q}_{12}^a\bar{q}_{12}^b})}
         \\\\
         C^{(ab)}=\beta\gamma(\bar{p}_{11}^{ab}-\bar{p}_{12}^{ab})%=\red{\beta\gamma(\sqrt{\bar{p}_{11}^a\bar{p}_{11}^b}-\sqrt{\bar{p}_{12}^a\bar{p}_{12}^b})}
         .&& 
    \end{array}
\end{equation}
where $\gamma=\lim\limits_{N\to\infty} K/N$, can be written as 
\begin{align}
\label{eq:streaming_apx}
    \frac{d\mathcal{A}^{\red{RS}}(\beta, H, \bm g,\boldsymbol h;t)}{dt}=  -\dfrac{\beta}{2}\SOMMA{a,b=1}{L}\SOMMA{\mu=1}{L}\bar{m}_\mu^a g_{ab}\bar{m}_\mu^{b}-\dfrac{\beta\gamma}{2}\SOMMA{\substack{a,b=1\\ a \neq b}}{L}\Big(\bar{p}_{11}^{ab}\bar{q}_{11}^{ab}-\bar{p}_{12}^{ab}\bar{q}_{12}^{ab}\Big)-\dfrac{\beta\gamma}{2}\SOMMA{a=1}{L}\bar{p}_{12}^{a}\Big(1-\bar{q}_{12}^{a}\Big).
\end{align}
Now we only need to compute the one-body term ($\mathcal{A}^{\red{RS}}(\beta, H, \bm g,\boldsymbol h;t=0)$). We start form \eqref{eq:interp_part_general} setting $t=0$
\begin{equation}
\begin{array}{lll}
     \mathcal{Z}_N(\beta, H,\bm g, \boldsymbol h;t=0)&=& \SOMMA{\{\bm\sigma^{1}\}}{}\cdots \SOMMA{\{\bm\sigma^{L}\}}{}\displaystyle\int\mathcal{D}(z)\exp\Bigg[\beta H \SOMMA{i=1}{N}\SOMMA{a=1}{L} h_i^a(t)\sigma_i^a+
     \\\\
     &&  + N \SOMMA{\mu=1}{L}\SOMMA{a=1}{L}\psi^{(a)}m_\mu^a+\SOMMA{\mu>L}{K}\tilde Y_\mu \SOMMA{a=1}{L}B^{(a)}z_{\mu}^a+\SOMMA{i=1}{N}Y_i \SOMMA{a=1}{L}A^{(a)}\sigma_i^a
     \\\\
     &&+\dfrac{1}{2}\SOMMA{i=1}{N} \SOMMA{\substack{a,b=1\\ a \neq b}}{L}C^{(ab)}\sigma_i^a\sigma_i^b+\dfrac{1}{2}\SOMMA{\mu>L}{K} \SOMMA{a,b=1}{L}\tilde{C}^{(ab)}z_{\mu}^az_{\mu}^b \Bigg]
\end{array}
\end{equation}
then using the definition \eqref{eq:def_stat_press} we can now  compute the one-body statistical pressure
\begin{align}
    \mathcal{A}^{\red{RS}}(\beta, H, \bm g,\boldsymbol h;t=0) = \lim_{N\to\infty} \frac{1}{N} \mathbb E \ln \mathcal{Z}_N(\beta,H,\boldsymbol g, \boldsymbol \xi, \boldsymbol h;t=0).
\end{align}
After some algebra we end up with
\begin{equation}
\label{eq:pressure_app_quad}
\begin{array}{lll}
     \mathcal{A}^{\red{RS}}(\beta, H, \bm g,\boldsymbol h;t=0)&=& \mathbb{E}_{\bm\xi, x}\log\Bigg\{\SOMMA{\{\bm\sigma^{a}\}}{}\exp\Bigg(\SOMMA{a=1}{L}\left[\SOMMA{\mu=1}{L}\beta\left(\bar{m}_\mu^a-\lambda\SOMMA{\substack{b=1\\ b \neq a}}{L}\bar{m}_\mu^{b}\right)\xi^\mu+\beta H  h^a(t)+x \sqrt{\beta \gamma \bar{p}_{12}^{a}}\right]\sigma^{a}
     \\\\
     &&
     +\SOMMA{\substack{b=1\\ b \neq a}}{L}\beta\gamma(\bar{p}_{11}^{ab}-\bar{p}_{12}^{ab})\sigma^{a}\sigma^{(b)}\Bigg)\Bigg\}
     \\\\
     &&  -\dfrac{\gamma}{2}\log\Big[\mathrm{det}\, \mathcal{G}\Big]+\dfrac{\beta\gamma}{2}\SOMMA{a,b=1}{L} \sqrt{\bar{q}_{12}^{a}} (\mathcal{G}^{-1}\,)_{ab}\, \sqrt{\bar{q}_{12}^{b}}
     % \\\\
     % &&
     % -\dfrac{\beta}{2}\SOMMA{a,b=1}{L}\SOMMA{\mu=1}{L}\bar{m}_\mu^ag_{ab}\bar{m}_\mu^{b}-\dfrac{\beta\gamma}{2}\SOMMA{a\neq b}{}\Big(\bar{p}_{11}^{ab}\bar{q}_{11}^{ab}-\bar{p}_{12}^{ab}\bar{q}_{12}^{ab}\Big)-\dfrac{\beta\gamma}{2}\SOMMA{a}{}\bar{p}_{12}^{a}\Big(1-\bar{q}_{12}^{a}\Big)
\end{array}
\end{equation}
where we have set
\begin{equation}
\begin{array}{lll}
     \mathcal{G}_{ab} &=& (g^{-1})_{ab} - \delta_{ab} C^{(a)} - (1-\delta_{ab})\tilde{C}^{(ab)} .
\end{array}
\end{equation}
Exploiting once more a $L-$dimensional multivariate Gaussian transform, we can linearize the last term of the argument of the exponential function in \eqref{eq:pressure_app_quad} and explicitly perform the sum over $\{\sigma^{a}\}$, getting the one-body statistical pressure
\begin{equation}
\begin{array}{lll}
     \mathcal{A}^{\red{RS}}(\beta, H, \bm g,\boldsymbol h;t=0)&=&-\dfrac{\beta\gamma}{2}+L\log{2} +\SOMMA{a=1}{L}\mathbb{E}_{\bm\xi, x}\cosh\Bigg(\left[\SOMMA{\mu=1}{L}\beta\xi^\mu\SOMMA{b=1}{L}g_{ab}\bar{m}_\mu^{b}+\beta H h^{a}(t)+x \sqrt{\beta \gamma \bar{p}_{12}^{a}}\right]\Bigg)
     \\\\
     &&  -\dfrac{1}{2}\log\Big[\mathrm{det}\mathcal{V}\Big]-\dfrac{\gamma}{2}\log\Big[\mathrm{det}\, \mathcal{G}\Big]+\dfrac{\beta\gamma}{2}\SOMMA{a,b=1}{L} \sqrt{\bar{q}_{12}^{a}} (\mathcal{G}^{-1}\,)_{ab}\, \sqrt{\bar{q}_{12}^{b}}
\end{array}
\end{equation}
where
\begin{equation}
\begin{array}{lll}
   \displaystyle\int \mathcal{D}(\tau)=\displaystyle\int \prod\limits_{b=1}^{L}\dfrac{d\tau_a d\tau_b}{2\pi } \exp\left(-\dfrac{1}{2}\SOMMA{b=1}{L}\tau_a(\mathcal{V}^{-1})_{ab}\,\tau_b\right)
\end{array}
\label{eq:final_one_body}
\end{equation}
and  $\mathcal{V}_{ab}=\delta_{ab}+(1-\delta_{ab})(\bar{p}_{11}^{ab}-\bar{p}_{12}^{ab})$.
\\
Finally, put Eqs.\eqref{eq:streaming_apx} and \eqref{eq:final_one_body} back in \eqref{eq:calc_theorem} we end up with the final expression of the statistical pressure of our model
\begin{equation}
\begin{array}{lll}
     \mathcal{A}^{\red{RS}}(\beta, H, \bm g,\boldsymbol h)&=&-\dfrac{\beta\gamma}{2}+L\log{2} +\SOMMA{a=1}{L}\mathbb{E}_{\bm\xi, x}\cosh\Bigg(\left[\SOMMA{\mu=1}{L}\beta\xi^\mu\SOMMA{b=1}{L}g_{ab}\bar{m}_\mu^{b}+\beta H h^{a}(t)+x \sqrt{\beta \gamma \bar{p}_{12}^{a}}\right]\Bigg)
     \\\\
     &&  -\dfrac{1}{2}\log\Big[\mathrm{det}\mathcal{V}\Big]-\dfrac{\gamma}{2}\log\Big[\mathrm{det}\, \mathcal{G}\Big]+\dfrac{\beta\gamma}{2}\SOMMA{a,b=1}{L} \sqrt{\bar{q}_{12}^{a}} (\mathcal{G}^{-1}\,)_{ab}\, \sqrt{\bar{q}_{12}^{b}}
     \\\\
     &&
     -\dfrac{\beta}{2}\SOMMA{a,b=1}{L}\SOMMA{\mu=1}{L}\bar{m}_\mu^ag_{ab}\bar{m}_\mu^{b}-\dfrac{\beta\gamma}{2}\SOMMA{\substack{a,b=1\\ a \neq b}}{L}\Big(\bar{p}_{11}^{ab}\bar{q}_{11}^{ab}-\bar{p}_{12}^{ab}\bar{q}_{12}^{ab}\Big)-\dfrac{\beta\gamma}{2}\SOMMA{a=1}{L}\bar{p}_{12}^{a}\Big(1-\bar{q}_{12}^{a}\Big).
\end{array}
\label{eq:semifinal_pressure}
\end{equation}
\red{Since stationary configurations correspond to those values of the order parameters that maximize the statistical pressure (or equivalently, minimize the free energy) of the system, and in this analysis we are only interested in the values of the order parameters that are saddle points of the free energy \eqref{eq:semifinal_pressure}, the previous expression can be further simplified by observing that its extremization with respect to $\bar{q}_{11}^{ab}$ and $\bar{q}_{12}^{ab}$ leads to the following relations:}
\begin{equation}
    \bar{q}_{11}^{ab}=\bar{q}_{12}^{ab} \;\;\;\;\;\bar{p}_{11}^{ab}=\bar{p}_{12}^{ab}
\end{equation}
which allow us to simplify \eqref{eq:semifinal_pressure} as
\begin{equation}
\label{eq:final_pressure}
\begin{array}{lll}
     % \mathcal{A}_N(\beta)&=& L\log 2+\SOMMA{a=1}{L}\mathbb{E}_{\bm\xi, x}\log\Bigg\{\mathbb{E}_{\tau_a}\cosh\Bigg[\SOMMA{\mu=1}{L}\beta\xi^\mu \SOMMA{b=1}{L}g_{ab}\bar{m}_\mu^{b}+m_\mu^b\beta h^{(a)}(t)+x \sqrt{\beta \gamma \bar{p}_{12}^{a}}+\tau_a\sqrt{\beta\gamma}\Bigg]\Bigg\}
     % \\\\
     % && -\dfrac{\gamma}{2}\log\Big[\mathrm{det}\mathcal{G}\Big]+\dfrac{\beta\gamma}{2}\SOMMA{a,b=1}{L} \sqrt{\bar{q}_{12}^{a}} (\mathcal{G}^{-1}\,)_{ab}\, \sqrt{\bar{q}_{12}^{b}}
     % \\\\
     % &&
     % -\dfrac{\beta}{2}\SOMMA{a,b=1}{L}\SOMMA{\mu=1}{L}\bar{m}_\mu^ag_{ab}\bar{m}_\mu^{b}-\dfrac{\beta\gamma}{2}\SOMMA{a}{}\bar{p}_{12}^{a}\Big(1-\bar{q}_{12}^{a}\Big)
     \mathcal{A}^{\red{RS}}(\beta, H, \bm g,\boldsymbol h)&=& L\log 2+\SOMMA{a=1}{L}\mathbb{E}_{\bm\xi, x}\log\Bigg\{\cosh\Bigg[\SOMMA{\mu=1}{L}\beta\xi^\mu \SOMMA{b=1}{L}g_{ab}\bar{m}_\mu^{b}+\beta H h^{a}(t)+x \sqrt{\beta \gamma \bar{p}_{12}^{a}}\Bigg]\Bigg\}
     \\\\
     && -\dfrac{\gamma}{2}\log\Big[\mathrm{det}\mathcal{G}\Big]+\dfrac{\beta\gamma}{2}\SOMMA{a,b=1}{L} \sqrt{\bar{q}_{12}^{a}} (\mathcal{G}^{-1}\,)_{ab}\, \sqrt{\bar{q}_{12}^{b}}
     \\\\
     &&
     -\dfrac{\beta}{2}\SOMMA{a,b=1}{L}\SOMMA{\mu=1}{L}\bar{m}_\mu^ag_{ab}\bar{m}_\mu^{b}-\dfrac{\beta\gamma}{2}\SOMMA{a=1}{L}\bar{p}_{12}^{a}\Big(1-\bar{q}_{12}^{a}\Big)
\end{array}
\end{equation}
where
\begin{equation}
   \mathcal{G}_{ab} =  \Big(1-\beta(1-\bar{q}_{12}^{a})\Big)\delta_{ab}-\lambda(1-\delta_{ab}).
\end{equation}
Where the order parameters must fullified the following self consistency equations
\begin{equation}
\begin{array}{lll}
     \bar{m}_{\nu}^{a}&=&  \mathbb{E}_{\bm\xi, x}\left\{\tanh\Bigg[\SOMMA{\mu=1}{L}\beta \xi^\mu\SOMMA{b=1}{L}g_{ab}\bar{m}_\mu^{b}+\beta H h^{a}(t)+x \sqrt{\beta \gamma \bar{p}_{12}^{a}}\Bigg]\xi^\nu\right\},
\end{array}
\end{equation}
\begin{equation}
\begin{array}{lll}
     \bar{q}_{12}^{a}&=& \mathbb{E}_{\bm\xi, x}\left\{\tanh^2\Bigg[\SOMMA{\mu=1}{L}\beta \xi^\mu\SOMMA{b=1}{L}g_{ab}\bar{m}_\mu^{b}+\beta H h^{a}(t)+x \sqrt{\beta \gamma \bar{p}_{12}^{a}}\Bigg]\right\},
\end{array}
\end{equation}
\begin{equation}
    \begin{array}{lll}
         \bar{p}_{12}^{c}&=& \dfrac{1}{\beta}\dfrac{\partial_{\bar{q}_{12}^{c}}\Big[\mathrm{det}\mathcal{G}\Big]}{\mathrm{det}\mathcal{G}}-\partial_{\bar{q}_{12}^{c}}\left[\SOMMA{a,b=1}{L} \sqrt{\bar{q}_{12}^{a}} \;(\mathcal{G}^{-1}\,)_{ab}\, \sqrt{\bar{q}_{12}^{b}}\right],
    \end{array}
\end{equation}
\red{which are obtained by the extremization of the \eqref{eq:final_pressure} with respect to $\bar{m}_{\nu}^{a}, \bar{q}_{12}^{a}$ and $\bar{p}_{12}^{c}$}.
\newline
\red{In this work we will specialize on the low-load regime, i.e. $\gamma =0$,  where the RS assumption is exact, much as like the standard Hopfield model, e.g., see \cite{Bovier}. On the other hand, in the high-load regime $\gamma \in \mathbb{R}^+$,  Replica-Symmetry-Breaking (RSB) phenomena are expected to emerge and their onset can, for instance, be addressed by determining the so-called de Almeida-Thouless line, e.g., see \cite{AT-Linda}, that traces -in the space of the control parameters- the boundaries of the stability of the RS solution.}

\setcounter{equation}{0}
\renewcommand\theequation{B.\arabic{equation}}

\section{Low-load self-consistency equations for $L=3$}
\label{subsec:explicit_L3}
%%%%%%%%%%
%%%%%%%%%%
%We now inspect the noisy scenario, where Eq.~\eqref{eq:evolv} is %revised as 
%\begin{equation}\label{eq:magn_evolv_beta}
%\sigma^{a}_i(t+1) = \textrm{sign} [\tilde{h}_i^{a}(t) + \textrm{atanh}%(\zeta_i^a(t))], 
%\end{equation}
%with $\zeta \sim \mathcal U(0,1)$. Provided that this rule is applied %long enough, the system will reach an equilibrium state where %configurations are sampled from the Boltzmann-Gibbs distribution %$P(\boldsymbol \sigma) \propto e^{-\beta \mathcal H(\boldsymbol %\sigma)}$ and the overall behavior of the system can be described by %the expected magnetizations $\{  \bar{m}_{\mu}^a \}_{\mu=1,...,K}$ and %overlaps $\bar{q}_{12}^{a}$.
%
%Resuming \eqref{eq:self}, we set $L=3$ and 
In this appendix we consider the general self-consistency equations \eqref{eq:self}, setting $L=3$ and
\begin{equation}
    \bm h^a(t) = \mathrm{sign}\left(\bm \xi^1+ \bm \xi^2+ \bm \xi^3\right)\;\;\;\;\mathrm{for}\;a=1,2,3,
\end{equation}
and look for numerically more-friendly expressions. First, it is convenient to define
\begin{equation}
    {\bm{\bar m}_{\mu}} = \left(\bar{m}_{\mu}^{1},\bar{m}_{\mu}^{2},\bar{m}_{\mu}^{3}\right),
\end{equation}
also
\begin{equation}
    \begin{array}{lll}
         \mathcal{T}_{++}^{a}(\bar{\bm{m}}_1,\bar{\bm{m}}_2,\bar{\bm{m}}_3) = \tanh\left[\beta\SOMMA{b=1}{3}g_{ab}\left(\bar{m}_1^b+\bar{m}_2^b+\bar{m}_3^b\right)+\beta H +x \sqrt{\beta \gamma \bar{p}_{12}^{a}}\right]
         \\\\
         \mathcal{T}_{+-}^{a}(\bar{\bm{m}}_1,\bar{\bm{m}}_2,\bar{\bm{m}}_3) = \tanh\left[\beta\SOMMA{b=1}{3}g_{ab}\left(\bar{m}_1^b+\bar{m}_2^b-\bar{m}_3^b\right)+\beta H +x \sqrt{\beta \gamma \bar{p}_{12}^{a}}\right]
         \\\\
         \mathcal{T}_{-+}^{a}(\bar{\bm{m}}_1,\bar{\bm{m}}_2,\bar{\bm{m}}_3) = \tanh\left[\beta\SOMMA{b=1}{3}g_{ab}\left(\bar{m}_1^b-\bar{m}_2^b+\bar{m}_3^b\right)+\beta H +x \sqrt{\beta \gamma \bar{p}_{12}^{a}}\right]
         \\\\
         \mathcal{T}_{--}^{a}(\bar{\bm{m}}_1,\bar{\bm{m}}_2,\bar{\bm{m}}_3) = \tanh\left[\beta\SOMMA{b=1}{3}g_{ab}\left(\bar{m}_1^b-\bar{m}_2^b-\bar{m}_3^b\right)-\beta H +x \sqrt{\beta \gamma \bar{p}_{12}^{a}}\right]
    \end{array}
    \label{eq:T_tilde}
\end{equation}
and
\begin{equation}
    \mathrm{det}\tilde{\mathcal{G}} = 1-\SOMMA{a=1}{3}d^{a}+(1-\lambda^2)\Big[d^{1}(d^{2}+d^{3})+d^{2}d^{3}\Big]-\mathrm{det}g\prod\limits_{a=1}^{3}d^{a}
\end{equation}
where we posed $d^{i}=\beta(1-\bar{q}_{12}^{i})$ and
\begin{equation}
    \begin{array}{lll}
         \bar{p}_{12}^{1}&=& \lambda\sqrt{\dfrac{\bar{q}_{12}^{3}}{\bar{q}_{12}^{1}}}\dfrac{1-(1+\lambda)d^{3}}{\mathrm{det}\tilde{\mathcal{G}}}+\lambda\sqrt{\dfrac{\bar{q}_{12}^{2}}{\bar{q}_{12}^{1}}}\dfrac{1-(1+\lambda)d^{2}}{\mathrm{det}\tilde{\mathcal{G}}}
         \\\\
         &&-\dfrac{\beta}{\mathrm{det}\tilde{\mathcal{G}}}\left\{\bar{q}_{12}^{2}[1-\lambda^2-(1+\lambda^2)(1-2\lambda)d^{3}]+\bar{q}_{12}^{3}[1-\lambda^2-(1+\lambda^2)(1-2\lambda)d^{2}]-2\lambda\sqrt{\bar{q}_{12}^{2}\bar{q}_{12}^{3}}\right\}
         \\\\
         &&+\dfrac{\beta}{[\mathrm{det}\tilde{\mathcal{G}}]^2}\Bigg[1-(1-\lambda^2)\SOMMA{i=2}{3}d^{i}+(1+\lambda^2)(1-2\lambda)\prod\limits_{i=2}^{3}d^{i}\Bigg]\SOMMA{c,b=1}{L} \sqrt{\bar{q}_{12}^{c}\bar{q}_{12}^{b}} \mathcal{M}_{cd}
    \end{array}
\end{equation}
being
\begin{equation}
\begin{array}{lll}
      \mathcal{M}=\begin{pmatrix}
          1-(1-\lambda^2)\SOMMA{i\neq 1}{3}d^{i}+\mathrm{det}g\prod\limits_{i\neq 1}^{3}d^{i} & -\lambda [1-(1+\lambda)d^{3}]&-\lambda [1-(1+\lambda)d^{2}]
          \\
          -\lambda [1-(1+\lambda)d^{3}]& 1-(1-\lambda^2)\SOMMA{i\neq 2}{3}d^{i}+\mathrm{det}g\prod\limits_{i\neq 2}^{3}d^{i} &-\lambda [1-(1+\lambda)d^{1}]
          \\
          -\lambda [1-(1+\lambda)d^{2}]& -\lambda [1-(1+\lambda)d^{1}]& 1-(1-\lambda^2)\SOMMA{i\neq 3}{3}d^{i}+\mathrm{det}g\prod\limits_{i\neq 3}^{3}d^{i}
      \end{pmatrix}.
\end{array}
\end{equation}
Then, defining
\begin{equation}
\begin{array}{lll}
     f_1^{a}(\bm x, \bm y, \bm z)&=&  \dfrac{1}{4}\mathbb{E}_{x}\left\{\mathcal{T}_{++}^{a}(\bm x, \bm y, \bm z) +\mathcal{T}_{+-}^{a}(\bm x, \bm y, \bm z)+\mathcal{T}_{-+}^{a}(\bm x, \bm y, \bm z)+\mathcal{T}_{--}^{a}(\bm x, \bm y, \bm z)\right\}\,,
\\\\
     f_2^{a}(\bm x, \bm y, \bm z)&=&  \dfrac{1}{4}\mathbb{E}_{x}\left\{[\mathcal{T}_{++}^{a}(\bm x, \bm y, \bm z)]^2 +[\mathcal{T}_{+-}^{a}(\bm x, \bm y, \bm z)]^2+[\mathcal{T}_{-+}^{a}(\bm x, \bm y, \bm z)]^2+[\mathcal{T}_{--}^{a}(\bm x, \bm y, \bm z)]^2\right\}
\end{array}
\label{eq:f_piccolo}
\end{equation}
we find
\begin{equation}\label{eq:self3}
\begin{array}{lllll}
     \bar{m}_{1}^{a}=  f_1^{a}(\bar{\bm{m}}_1,\bar{\bm{m}}_2,\bar{\bm{m}}_3), &&\bar{m}_{2}^{a}=  f_1^{a}(\bar{\bm{m}}_2,\bar{\bm{m}}_1,\bar{\bm{m}}_3), && 
\\\\
     \bar{m}_{3}^{a}=  f_1^{a}(\bar{\bm{m}}_3,\bar{\bm{m}}_2,\bar{\bm{m}}_1), &&\bar{q}_{12}^{a}=  f_2^{a}(\bar{\bm{m}}_1,\bar{\bm{m}}_2,\bar{\bm{m}}_3).
\end{array}
\end{equation}

Of course, when $\lambda =0$ we recover the self-consistency equations of three independent Hopfield models. 

Moreover, in the low-load regime ($\gamma =0$), we have 
\begin{equation}\label{eq:self3}
\begin{array}{lllll}
     \bar{m}_{1}^{a}=  f_1^{a}(\bar{\bm{m}}_1,\bar{\bm{m}}_2,\bar{\bm{m}}_3),
\\\\
    \bar{m}_{2}^{a}=  f_1^{a}(\bar{\bm{m}}_2,\bar{\bm{m}}_1,\bar{\bm{m}}_3), 
\\\\
     \bar{m}_{3}^{a}=  f_1^{a}(\bar{\bm{m}}_3,\bar{\bm{m}}_2,\bar{\bm{m}}_1).
\end{array}
\end{equation}
where $f_1^{a}(\bm x, \bm y, \bm z)$ is defined in the first row of \eqref{eq:f_piccolo} and \eqref{eq:T_tilde} simplify to
\begin{equation}
    \begin{array}{lll}
         \mathcal{T}_{++}^{a}(\bar{\bm{m}}_1,\bar{\bm{m}}_2,\bar{\bm{m}}_3) = \tanh\left[\beta\SOMMA{b=1}{3}g_{ab}\left(\bar{m}_1^b+\bar{m}_2^b+\bar{m}_3^b\right)+\beta H \right]\,,
         \\\\
         \mathcal{T}_{+-}^{a}(\bar{\bm{m}}_1,\bar{\bm{m}}_2,\bar{\bm{m}}_3) = \tanh\left[\beta\SOMMA{b=1}{3}g_{ab}\left(\bar{m}_1^b+\bar{m}_2^b-\bar{m}_3^b\right)+\beta H \right]\,,
         \\\\
         \mathcal{T}_{-+}^{a}(\bar{\bm{m}}_1,\bar{\bm{m}}_2,\bar{\bm{m}}_3) = \tanh\left[\beta\SOMMA{b=1}{3}g_{ab}\left(\bar{m}_1^b-\bar{m}_2^b+\bar{m}_3^b\right)+\beta H \right]\,,
         \\\\
         \mathcal{T}_{--}^{a}(\bar{\bm{m}}_1,\bar{\bm{m}}_2,\bar{\bm{m}}_3) = \tanh\left[\beta\SOMMA{b=1}{3}g_{ab}\left(\bar{m}_1^b-\bar{m}_2^b-\bar{m}_3^b\right)-\beta H \right]\,.
    \end{array}
\end{equation}

\setcounter{equation}{0}
\renewcommand\theequation{C.\arabic{equation}}
\section{Calculations for the stability analysis in the noiseless, high-load regime} \label{sec:consistency}

Let us start our inspection with the state $\boldsymbol \sigma^{(1,2,3)} = (\boldsymbol \xi^1, \boldsymbol \xi^2, \boldsymbol \xi^3)$. 
This is our target configuration, whose magnetization is $m_{\mu}^a = \delta_{\mu a}$ for $a=1,...,3$ (apart from vanishing corrections in the thermodynamic limit). This configuration minimizes the first contribution in the cost function \eqref{eq:HamHam}, whose value can be estimated in the large size limit (we exploit the Rademacher nature of pattern entries and the central limit theorem, c.l.t.) to get 
\begin{equation} \label{eq:HHH0}
\frac{\mathcal H (\boldsymbol \sigma^{(1,2,3)})}{N} \underset{c.l.t.}{\sim} -3(1+\gamma) - \frac{3}{2}H+x\frac{\mathcal{C}^{(1,2,3)}}{\sqrt{N}}, 
\end{equation}
where we dropped the dependence on $\lambda, \boldsymbol \xi, H, \boldsymbol h$ to lighten the notation, $x\sim\mathcal{N}(0,1)$ and $\mathcal{C}^{(1,2,3)}$ is a constant depending only on $\gamma$, $H$ and $\lambda$. Notice that, by increasing $H$ and $\gamma$, the configuration $\boldsymbol \sigma^{(1,2,3)}$ gets energetically more favorable.
To check the consistency of these configurations we take $\boldsymbol \sigma^{(1,2,3)}$ as initial state, then, following \eqref{eq:magn_evolv}, we derive the \emph{next-step magnetization}, that is the magnetization corresponding to the configuration after one time step. In the thermodynamic limit this reads as  
 \begin{equation}
 m_1^{1}(t=1)=m_2^{2}(t=1)=m_3^{3}(t=1)=\mathrm{erf}\left[\dfrac{2+H}{\sqrt{2\left(4\gamma +8\lambda^2(1+\gamma)-8\lambda H +3 H^2\right)}}\right]. 
 \label{eq:erf_123}
 \end{equation}
As long as $\gamma$, $\lambda$, and $H$ are simultaneously sufficiently small, the r.h.s. coincides with $ m_1^{1}(t=0)=m_2^{2}(t=0)=m_3^{3}(t=0)=1$, thus, under these conditions, this configuration is a fixed point.\\
As expected, in the limit $H,\lambda \to 0$, \eqref{eq:erf_123} recovers the expression for the next-step magnetization of three independent Hopfield models, each initialized with the respective initial condition $\boldsymbol \sigma^{a} = \boldsymbol \xi^a$, $a=1,2,3$.

Let us now consider the configuration $\bm \sigma^{(1,1,1)}  \equiv(\bm\xi^1,\bm\xi^1,\bm\xi^1)$, which corresponds to the pure retrieval in a standard Hopfield model and minimizes the first contribution in the cost function \eqref{eq:HamHam}. Its intensive energy in the large-size limit is
    \begin{equation} \label{eq:HHH}
    \frac{\mathcal H (\boldsymbol \sigma^{(1,1,1)})}{N} \underset{c.l.t.}{\sim}-3(1-\lambda)(1+\gamma)-\frac{3}{2}H +x\frac{\mathcal{C}^{(1,1,1)}}{\sqrt{N}}.
    \end{equation}
    As expected, when $\lambda$ is increased, this configuration makes the coupling between layers more  frustrated, consequently, its energy grows and the related stability of the solution gets impaired; if $\lambda=0$ the above energy recovers the previous one for $\bm \sigma^{(1,2,3)}$.
    The next-step magnetization in the thermodynamic limit is
    \begin{equation}\label{eq:erf_111}
    m_1^{1}(t=1)=m_2^{2}(t=1)=m_3^{3}(t=1)=\mathrm{erf}\left[\dfrac{2(1-2\lambda)+H}{\sqrt{2\left(4\gamma(1-2\lambda)^2+3 H^2\right)}}\right].
    \end{equation}
    Notice that, for relatively small fields $H$ and for relatively small couplings $\lambda$, consistency can be recovered.

Next, we consider the staggered configuration $\bm\sigma^{(1,1,1')} \equiv(\bm\xi^1,\bm\xi^1,-\bm\xi^1)$, which  minimizes both the first and the third contribution of the cost function \eqref{eq:HamHam}.
    The intensive energy is 
    \begin{equation}
        \frac{\mathcal H (\boldsymbol \sigma^{(1,1,1')})}{N}\underset{c.l.t.}{\sim}-(3+\lambda)(1+\gamma)-\dfrac{H}{2}+x\frac{\mathcal{C}^{(1,1,1')}}{\sqrt{N}}.
        \label{eq:HHH11_1}
    \end{equation}
    By comparing this expression with \eqref{eq:HHH0}, \eqref{eq:HHH} and the following \eqref{eq:HHH1}, we see that, when $H=0$ and $\lambda \neq 0$, this state is the one with the lowest energy among those considered here, in fact, this configuration favors all the intra-layer interactions and partially favours inter-layer interactions. However, by comparing this energy with the one obtained for $\bm\sigma^{(1,2,3)}$, we see that there exists a range of values for the parameters $H \neq 0$ and $\lambda$, such that the energy of this state is larger and therefore energetically less convenient.\\
    In the thermodynamic limit, the next-step magnetization is the same for layers $a=1,2$, that is, 
    \begin{equation} \label{eq:erf_11a}
   m_1^{1}(t=1)=m_1^{2}(t=1)=\mathrm{erf}\left[\dfrac{2+H}{\sqrt{2\left(4\gamma+3H^2\right)}}\right],
    \end{equation}
   while for the third layer 
    \begin{equation} \label{eq:erf_11b}
   m_1^{3}(t=1)=- \mathrm{erf}\left[\dfrac{2+4\lambda-H}{\sqrt{2\left[4\gamma(1+2\lambda)^2+3H^2\right]}}\right].
    \end{equation}
    Notice that, if $H=0$ and $\gamma\ll 1$, $m_1^{1}(t=1)=m_1^{2}(t=1)\approx 1$ and their expression recovers the one of a pure state in a standard Hopfield model. 
    Further, if $\lambda\neq 0$, $|m_1^{3}(t=1)|$ is as well close to 1 and it is enhanced by $\lambda$ (in fact, the denominator is always smaller than $1$ if $0<\lambda<1/2$).

 Finally, we focus on $\bm\sigma^{(h)} \equiv(\bm h,\bm h,\bm h)$.
 This state corresponds to the input mixture, repeated over all the layers. In the large $N$ limit the related intensive energy is 
 \begin{equation} \label{eq:HHH1}
 \frac{\mathcal H (\boldsymbol \sigma^{(h)})}{N}\underset{c.l.t.}{\sim}-3(1-\lambda)\left(\dfrac{3}{4}+\gamma\right)-3H+x\frac{\mathcal{C}^{(h)}}{\sqrt{N}},
  \end{equation}
 which, as expected, decreases (increases) monotonically with $H$ (with $\lambda$).\\
 Further, recalling $\boldsymbol h = \textrm{sign}(\boldsymbol \xi^1 + \boldsymbol \xi^2 + \boldsymbol \xi^3)$, the next-step magnetization is the same for all layers and reads as 
     \begin{equation} \label{eq:erf_h}
     \dfrac{1}{N}\SOMMA{i=1}{N}h_i  \sigma_i^{a}(t=1) \underset{N \to \infty}{=}\mathrm{erf}\left[\dfrac{\frac{3}{4}(1-2\lambda)+H}{\sqrt{2[\left(\gamma+\frac{3}{16}\right)(1-2\lambda)^2] }}\right]\;\;\;\;\mathrm{with}\;\;a=1,\cdots,3.
       \end{equation}
    Note that the invariance of this configuration is improved as the field $H$ increases. 
    %Moreover for $\gamma\ll 1$, this configuration is a fixed point for any possible value of $\lambda$ as long as $H>0$.

\setcounter{equation}{0}
\renewcommand\theequation{D.\arabic{equation}}
\section{\red{Spectrum of the free-energy Hessian in the low-load regime}}\label{sec:spectrum}

We resume the second-order derivative \eqref{eq:hessian} of $f^{RS}$
\begin{align}\label{eq:hessian}
    D_{\mu\nu}^{ab} = g_{ab} \delta_{\mu\nu} - \beta \SOMMA{c=1}{L} g_{cb}g_{ca} \mathbb{E}_{\bm\xi}\left\{ \xi^\mu\xi^\nu \left[ 1-\tanh^2\lr{\beta \sum_{\rho=1}^{L} \xi^\rho \SOMMA{d}{L} g_{cd} \overline m_\rho^d} \right] \right\}
\end{align}
and in this expression we recognize \red{(when $\mu = \nu$)} the overlap $\overline q^a_{12}$ between the $(1,2)$ replicas in the same layer $a$, that is 
\begin{equation}
    \overline q^a_{12} = \mathbb{E}_{\bm\xi}\left\{ \tanh^2\lr{\beta \sum_{\rho=1}^L \xi^\rho \sum_{d=1}^L g_{ad} \overline m_\rho^d}\right\},
    \end{equation}
  \red{obtained by suitably simplifying \eqref{eq:self}, and (when $\mu \neq \nu$)} the quantity $Q^{\mu\nu}_a$ defined as
    \begin{equation}
    Q^{\mu\nu}_a = \mathbb{E}_{\bm\xi}\left\{ \xi^\mu\xi^\nu\tanh^2\lr{\beta \sum_{\rho=1}^L \xi^\rho \sum_{d=1}^L g_{ad} \overline m_\rho^d}\right\}.
\end{equation}
Thus, we can recast the diagonal entries ($a=b$) of the Hessian matrix $D_{\mu\nu}^{aa}$ as
\begin{align}
\nonumber
    D_{\mu\nu}^{aa} = \delta_{\mu\nu} \left[1-\beta \lr{1-\overline q^a_{12}}+ \lambda^2 \SOMMA{\substack{c=1\\ c \neq a}}{L}(1-\overline q^c_{12})\right] + (1-\delta_{\mu\nu})\beta\left[ Q^{\mu\nu}_a + \lambda^2 \SOMMA{\substack{c=1\\ c \neq a}}{L} Q^{\mu\nu}_c \right],
\end{align}
and the off-diagonal entries ($a\neq b$) as
\begin{align}
\nonumber
    D_{\mu\nu}^{ab} = \delta_{\mu\nu} \lambda \left[ -1 + \beta (2-\overline q^a_{12}-\overline q^b_{12}) + \lambda \SOMMA{\substack{c=1\\ c \neq a,b}}{L}  (1-\overline q^c_{12})\right] +  (1-\delta_{\mu\nu})\beta\lambda \left[ - \lr{Q^{\mu\nu}_a + Q^{\mu\nu}_b} + \lambda \SOMMA{\substack{c=1\\ c \neq a,b}}{L} Q^{\mu\nu}_c \right].
\end{align}

Notice that, for $\mu=\nu$, $Q_a=\overline q^a_{12}$, while for $\mu\neq \nu$, $Q^{\mu\nu}_a$ is independent of the indices $\mu,\nu$ and it can be simply written as $Q_a=\mathbb{E}_{\bm\xi}\left\{ \xi^1\xi^2 \tanh^2\lr{\beta \sum_{\rho=1}^{L} \xi^\rho \sum_{d=1}^L g_{ad} m_{\rho d}}\right\}$.\\
Hence, for $a=b$ and $\mu,\nu\leq L$, the eigenvalues of $D_{\mu\nu}^{aa}$ with the related multiplicities read as
\begin{align}
    &t_1 = 1-\beta(1-\overline q^a_{12})-\beta\lambda^2 \SOMMA{\substack{c=1\\ c \neq a}}{L}  (1-\overline q^c_{12}), ~~~~~~~  \textrm{mult.} =K-L\\
    &t_2 = t_1 + (L-1) \beta Q_a + (L-1) \beta \lambda^2 \SOMMA{\substack{c=1\\ c \neq a}}{L}  Q_c,~~  \textrm{mult.} =1\\
    &t_3 = t_1 - (L-1) \beta Q_a - (L-1) \beta \lambda^2 \SOMMA{\substack{c=1\\ c \neq a}}{L}  Q_c, ~\,\, \textrm{mult.} =L-1.
\end{align}
%%%%
The eigenvalues can be computed numerically for different values of $\beta,\lambda$ and for the related estimates of the magnetisation matrices $\bar{\boldsymbol m}^{(1,2,3)}$ and $\bar{\boldsymbol m}^{(h)}$\red{, which, in turn, affect the value of $Q$. By stydying the sign of the smallest eigenvalue we can determine whether the solution is stable.}

\setcounter{equation}{0}
\renewcommand\theequation{E.\arabic{equation}}

\section{Details on computational experiments} \label{ssec:MC_details}

\red{In this section we report the technical details concerning the numerical solution of the self-consistency equations and the MC simulations, along with a discussion on the computation time scaling vs the system size. We start presenting the algorithm used to numerically solve a generic set of self-consistency equations, see Algorithm \ref{alg:SC}. To simplify the notation, we introduce the following definitions:}
\begin{equation}
    \begin{array}{lll}
         \bm M_\mu &=& (\bar{m}_\mu^1, \bar{m}_\mu^2, \cdots, \bar{m}_\mu^L)\;\;\;\mathrm{with}\;\;\mu=1,\cdots, L
         \\\\
         \bm Q &=& (\bar{q}_{12}^1, \bar{q}_{12}^2, \cdots, \bar{q}_{12}^L)
    \end{array}
    \label{eq:self_compresse_numerico}
\end{equation}
\red{and we consider a generic set of self-consistency equations of the form:}
$$
\bm M_\mu = f_{1,\mu}(\bm g, T, \gamma, \bm M_1, \cdots, \bm M_L, \bm Q)\;\;\;\mathrm{with}\;\;\mu=1,\cdots, L
$$
$$
\bm Q = f_2(\bm g, T, \gamma, \bm M_1, \cdots, \bm M_L, \bm Q)
$$
\begin{algorithm}[bt] 
\caption{Numerical solution of the self consistency equations \label{alg:SC}}
\KwIn{Load of the network $\gamma$, temperature $T$, starting magnetizations (${\bm M_1}^{start}, \cdots, {\bm M_L}^{start}$) and overlaps (${\bm Q}^{start}$), interaction matrix $\bm g$, maximum number of iterative steps $N_{iter}$, tolerance threshold $\delta^*$}
\KwOut{Value of the $L$ Mattis magnetization vectors (${\bm M_1}, \cdots, {\bm M_L}$) and overlaps (${\bm Q}$)}
\vspace*{0.1cm}
Set the starting points of the fixed point iterations:\\
$\bm M_1, \cdots, \bm M_L= ({\bm M_1}^{start}, \cdots, {\bm M_L}^{start})$;\\
$\bm Q={\bm Q}^{start}$\;
        \For{$iter$ in \( (1, \ldots, N_{iter}) \)}{
        Compute the r.h.s. of the self consistecy equations: 
        \\
        $\bm M^{new}_\mu= f_{1,\mu}(\bm g, T, \gamma, \bm M_1, \cdots, \bm M_L, \bm Q)\;\;\;\mathrm{with}\;\; \mu = 1, \cdots, L$;
        \\ 
        $\bm Q^{new}= f_2(\bm g, T, \gamma, \bm M_1, \cdots, \bm M_L, \bm Q)$\;
        Evaluate $\delta = \sqrt{\sum\limits_{\mu=1}^L|\bm{M}^{new}_\mu-\bm{M}_\mu|^2+|\bm{Q}^{new}-\bm{Q}|^2}$\;
            \If{$\delta< \delta^*$}{
        \textbf{break}}
        \Else{
        Compute the fixed point equations for the order parameters
        \\
        $\bm M_\mu = \frac{\bm M_\mu +\bm M^{new}_\mu}{2}\;\;\;\mathrm{with}\;\; \mu = 1, \cdots, L$;\\\vspace*{0.2cm}
        $\bm Q = \frac{\bm Q +\bm Q^{new}}{2}$}
        }
\end{algorithm}

\red{that are like those presented in \eqref{eq:self3}. We exploit the fixed-point iteration method to compute the value of the $L\times L$ Mattis magnetizations (first row of \eqref{eq:self_compresse_numerico}) and $L$ two-replica overlaps (second row of \eqref{eq:self_compresse_numerico}) for a fixed value of the temperature $\beta^{-1}$, the network load $\gamma$ and the interaction matrix $\bm g$.}
\red{
The algorithm is run for a fixed number of iterations  $N_{iter}$ or until a predefined tolerance threshold $\delta^*$ for the solution is reached — whichever comes first. For practical purposes, we set $N_{iter} = 10^3$ and $\delta^*=10^{-6}$, which represents a reasonable trade-off between convergence accuracy and computational time.}

\red{Now, we present the algorithm used to perform Monte Carlo simulations. We provide the pseudocode for the case of a generic number of layers $L$ both in the case of sequential and parallel updating, respectively Algorithm \ref{algo:2} and Algorithm \ref{algo:3}.}

\begin{algorithm}[tb]
\caption{MC Glauber dynamic: sequential updating \label{algo:2}}
\KwIn{Interaction matrix $\bm g$, patterns $\{\xi_i^\mu\}_{i=1, \cdots, N}^{\mu=1, \cdots, K}$, starting configurations $\bm\sigma^{1}(t=0), \cdots, \bm\sigma^{L}(t=0)$, external fields $\{h_i^a\}^{a=1,\cdots,L}_{i=1, \cdots, N}$, number of sequential dynamic steps $ N_s $, thermal noise $ T $}
\KwOut{Final neuronal configuration $\bm\sigma^{1}(t=N_s), \cdots, \bm\sigma^{L}(t=N_s) $}

Set $iter = 0$\;

\Repeat{$iter = N_s$}{
    Sample, with possible repetitions, $L$ random integers $n_1, \cdots, n_L$ uniformly in the set $ \{1, \ldots, N\} $\;
    Sample $L$ random variables $u_1,\cdots, u_L$ from a uniform distribution $\mathcal{U}(-1,1)$\;
    Randomly determine the order of layer updates by shuffling the vector $\bm A=[1, 2, 3, \cdots, L]$\;
    \For{$a$ in $\bm A$}{
    Update the $n_a$-th  neuron $\sigma_{n_a}^{a}$ according to \\
    $$\tilde{h}_{n_a}^{a}= \dfrac{1}{N}\SOMMA{b=1}{L}g_{ab}\SOMMA{\mu=1}{K}\SOMMA{i=1}{N}\xi_{n_a}^\mu\xi_i^\mu \sigma_i^b +H h^{a}_{n_a};$$
    \\
    $$\sigma_{n_a}^a = \text{sign} \left[\tanh\left(\frac{1}{T} \;\tilde{h}_{n_a}^{a}\right) + u_a\right]; $$
    }
    $ iter = iter + 1 $\;
}
\end{algorithm}
\red{To perform the MC simulation both in the case of parallel or sequential case, we start from the updating rules presented in Eq.~\eqref{eq:evolv}, which we report here for convenience.}
\begin{eqnarray}
\label{eq:evolv_MC}
\sigma^{a}_i(t+1) &=& \textrm{sign} [\tanh(\beta\tilde{h}_i^{a}(t)) +  u_i^{a}(t)] \\
      \tilde{h}_i^{a}(t) &=& \dfrac{1}{N}\SOMMA{b=1}{L}g_{ab}\SOMMA{\mu=1}{K}\SOMMA{j=1}{N}\xi_j^\mu \sigma_j^b(t) \xi_i^\mu+H h_i^a\label{eq:evolv_MC_h}
\end{eqnarray}
\red{where $t$ denotes the time step, $u_i^{a}(t)$ is an uniform random variable ranging in $[-1,+1]$.
These updating rules can be applied for a fixed number of iterations (e.g. to explore the one-step magnetizations), or until a stable configuration is reached (e.g. to inspect the stationary state magnetizations). The dynamics can be implemented either sequentially (Algorithm \ref{algo:2})— updating one layer, chosen randomly, a time and, within each layer, one neuron, always chosen randomly, at a time, recomputing the internal fields \eqref{eq:evolv_MC_h} after each individual neuronal update — or in parallel (Algorithm \ref{algo:3}), updating all layers and all neurons simultaneously using the internal fields from the previous step, and recomputing the internal fields only after all neurons in the network have been updated. 
For practical reasons, the number of iterations must be chosen differently to ensure convergence of both algorithms. In the sequential case, to be sure of the stability of our results we need that each neuron is updated at least once; therefore, the number of steps $N_s$ must be much larger than the total number of neurons in the network, i.e., $N_s \gg N \times L$. In contrast, in the parallel case, since all neurons are updated simultaneously, the number of steps required is significantly reduced, and it suffices that $N_p \gg L$.}
\begin{algorithm}[tb]
\caption{MC Glauber dynamic: parallel updating \label{algo:3}}
\KwIn{Interaction matrix $\bm g$, patterns $\{\xi_i^\mu\}_{i=1, \cdots, N}^{\mu=1, \cdots, K}$, starting configurations $\bm\Omega(t=0)=\Big(\bm\sigma^{1}(t=0), \cdots, \bm\sigma^{L}(t=0)\Big)$, external fields $\{h_i^a\}^{a=1,\cdots,L}_{i=1, \cdots, N}$, number of parallel dynamic steps $ N_p $, thermal noise $ T $}
\KwOut{Final neuronal configuration $\bm\Omega(t=N_p)=\Big(\bm\sigma^{1}(t=N_p), \cdots, \bm\sigma^{L}(t=N_p) \Big)$}

Set $iter = 0$\;

\Repeat{$iter = N_p$}{
    Sample a tensor of uniform distributed $\mathcal{U}(-1,1)$ random variables of dimension $N \times L$: $\bm U \in \mathbb{R}^{N\times L}$   \;
    Compute the $N \times L$ internal fields tensor
    $$\tilde{h}_i^{a}(t=iter)= \dfrac{1}{N}\SOMMA{b=1}{L}g_{ab}\SOMMA{\mu=1}{K}\SOMMA{j=1}{N}\xi_{i}^\mu\xi_j^\mu \sigma_j^{b}(t=iter-1) +H h^{a}_{i}\;\;\;\mathrm{with}\;\; ^{a=1,\cdots, L}_{i = 1,\cdots,N};$$
    Update the whole networks configurations
    $$\bm\Omega(t=iter) = \text{sign} \left[\tanh\left(\frac{1}{T} \;\tilde{\bm h}(t=iter)\right) + \bm U\right] $$
    $ iter = iter + 1 $\;
}
\end{algorithm}
%%%
\red{The code and datasets
supporting this work are publicly available at:} \href{https://github.com/AndreaAlessandrelli4/Coding_Lessons/tree/main/MC_simulation_Spurious}{MC\_simulation\_Spurious}.

\begin{figure}
    \centering
    \includegraphics[width=0.85\linewidth]{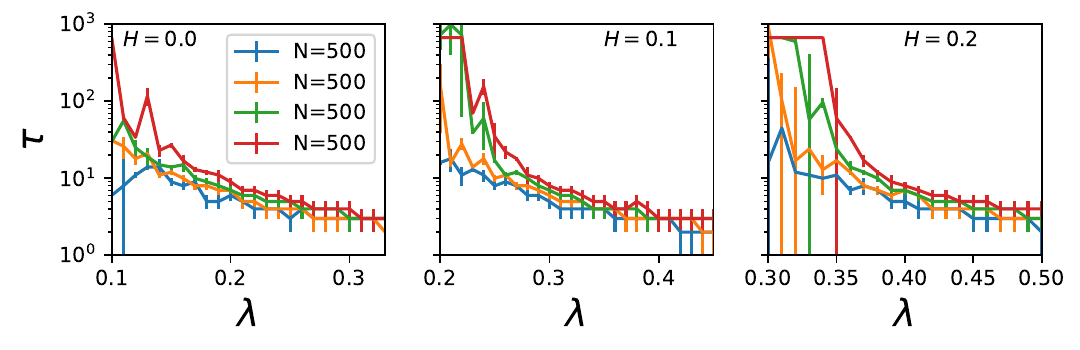}
    \includegraphics[width=0.85\linewidth]{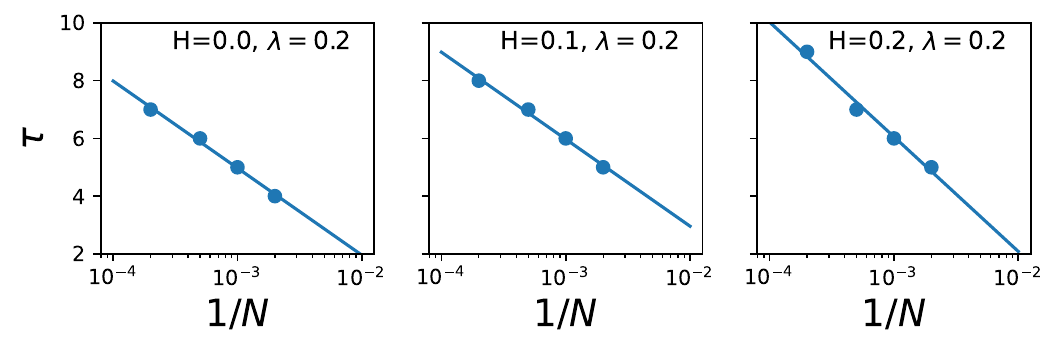}
    
    \caption{\red{Upper panel: the time $\tau$ of convergence  to the disentangled state has been measured by testing parallel MC dynamics for different sizes of the model, as reported in the legend, and for different values of the external field $H$. Lower panel: the finite-size- scaling for some choices of the parameters $H$ and $\lambda$ shows a logarithmic law for $\tau$ versus $N$.}}
    \label{fig:scaling}
\end{figure}

\red{We conclude this appendix with a brief discussion on how the computational time $\tau$ scales with the system size $N$. We recall that here, by ``computational time'' we mean the time required for the system, initialized in the configuration $\boldsymbol \sigma^{(h)}$, to reach the target configuration $\boldsymbol \sigma^{(1,2,3)}$ or more precisely, a configuration in which the magnetizations corresponding to the three mixed patterns exceed a chosen threshold, e.g., $m>0.95$, along the evolution mimicked by a MC simulation with parallel updating -- each update, involving $N$ neurons, counts one MC step, see also Algorithm \ref{algo:3}. The results obtained for several sizes $N$ are compared in Figure \ref{fig:scaling} (upper panels). The same experiment is repeated for different values of $\lambda$ and $H$ obtaining that, within the disentanglement region, $\tau$ scales logarithmically with $N$, as illustrated in Figure \ref{fig:scaling} (lower panels).}
%%%%

\begin{figure}[tb]
    \centering
    \includegraphics[width=12.5cm]{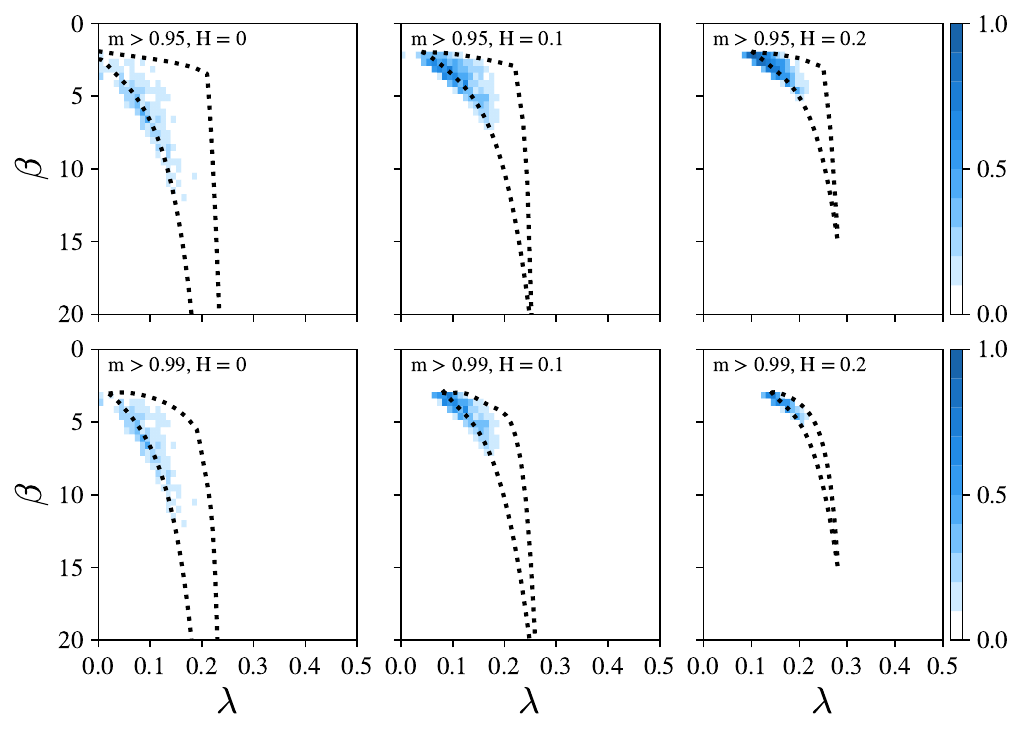}
    \caption{The region in the plane $(\beta, \lambda)$ where the five-layer model is expected to successfully disentangle mixtures of five patterns is depicted by solving the self-consistent equations (\ref{eq:self}) (dashed lines) and compared to MC simulations (color map), for three values of the external field: $H = 0.0$ (left column), $H = 0.1$ (middle column) and $H = 0.2$ (right column), and two different thresholds on the magnetizations $m > 0.95$ (upper row), $m > 0.99$ (lower row), in analogy to Figure \ref{fig:spurious-res}. The self-consistency equations have been solved in the $\gamma=0$ case, while the disentangling accuracy has been computed by averaging over $50$ statistically-independent MC runs, each with $N = 5000$ and $K = 5$. In each run the model is initialized in the $\boldsymbol \sigma^{(h)}$ configuration and let evolve up to convergence to a stationary state; the final magnetizations have been obtained by computing the overlap between the state of each layer and the five patterns $\boldsymbol \xi^1,..,\boldsymbol \xi^5$.}
    \label{fig:newMartino}
\end{figure}

\begin{figure}[tb]
    \centering
    \includegraphics[width=15cm]{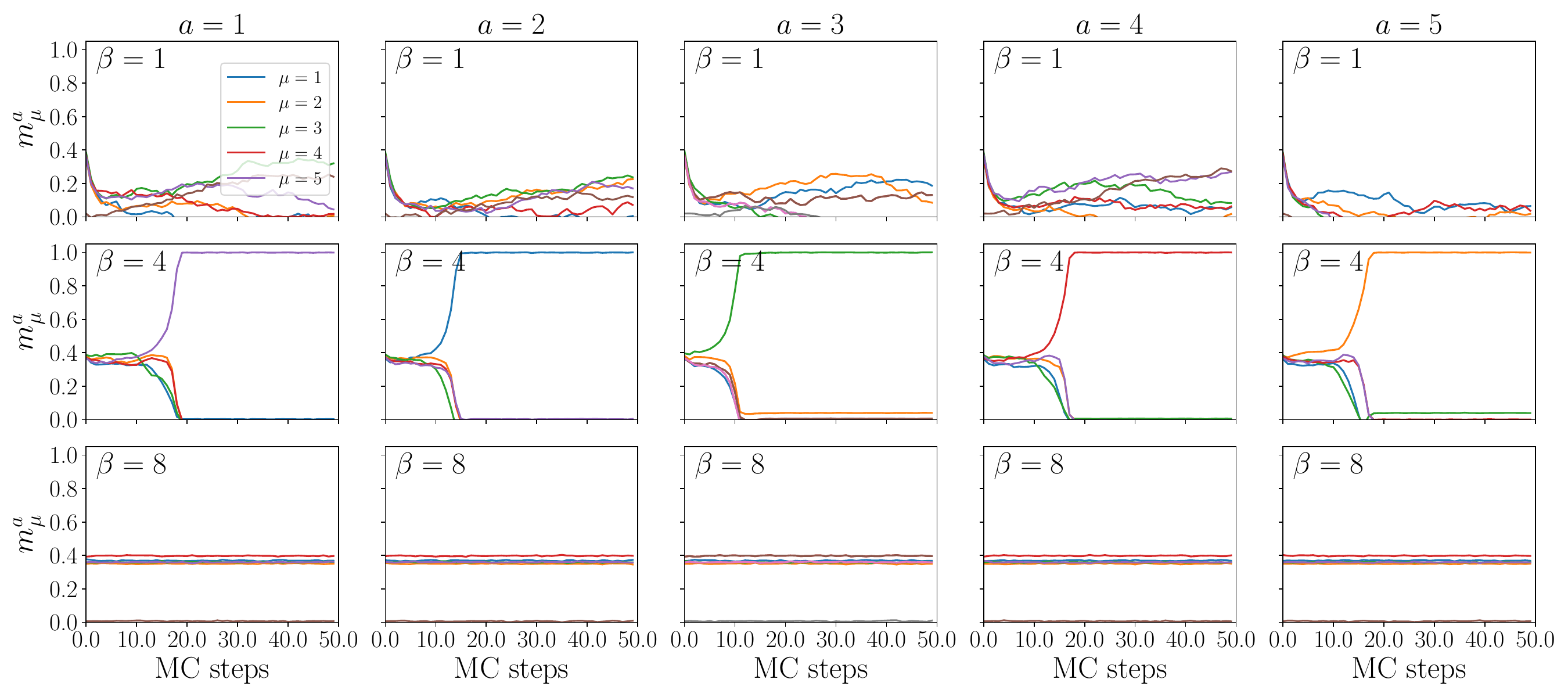}
    \caption{These plots show the evolution of the Mattis magnetizations $m_{\mu}^a$ for $\mu=1, ...,K$ (different labels correspond to different colors) and for $a=1, ..., 5$ (different layers correspond to different columns) versus the number of MC steps -- one MC step corresponds to $N$ random extractions of the index $i \in \{1, ..., N\}$ that identifies the neuron to be updated according to the rule \eqref{eq:evolv}. More precisely, here we set $N=5000$, $K=50$, $H=0.1$ and $\lambda=0.11$, while different values of $\beta$ are chosen: $\beta=1$ (upper row), $\beta=4$ (middle row), $\beta=8$ (lower row); in agreement with the findings presented in Figure~\ref{fig:newMartino}, the emerging behavior is, respectively, ergodic, disentangled, stuck in the spurious state. }
    \label{fig:GTAML5}
\end{figure}

\setcounter{equation}{0}
\renewcommand\theequation{F.\arabic{equation}}
\section{Checking the robustness of results: $L=5$} \label{sec:L5}

In this section we present some experiments run on a system made of $L=5$ layers to check the robustness of the results presented in the main text for $L=3$. 
In particular, following the procedure explained in Sec.~\ref{sec:ssr} , we handle the self-consistency equations \eqref{eq:self} in the low-load regime ($\gamma=0$) to outline a region in the plane ($\lambda, \beta$) where disentanglement can be accomplished. This region corresponds to the area in-between the dashed lines in Figure~\ref{fig:newMartino}. \red{We stress that we are requiring all five patterns to be unmixed and reconstructed with the same minimal level of quality. By comparing these results with those for the case $L=3$ (see Figure \ref{fig:spurious-res}), we observe that the disentanglement region is narrower.} 
Furthermore, we execute MC experiments to assess the network’s accuracy across different threshold values, as detailed in Sec.\ref{sec:MC}. The results obtained are consistent with those derived from the self-consistency equations and are also reported in Figure~\ref{fig:newMartino}.
\red{In particular, the region corresponding to a high success rate lies entirely within the theoretical bounds. Additionally, in this case, the external field appears to play a more significant role.}
Finally, these findings are corroborated in Figure~\ref{fig:GTAML5}, where we show the temporal evolution of the Mattis magnetization measured on the five layers for different choices of $\beta$.

\red{A more in-depth analysis of the scalability of network performance with respect to $L$ goes beyond the scope of the present work, which is primarily focused on highlighting non-trivial behaviors emerging from the interaction of coupled Hopfield networks. Algorithms specifically designed to address scalability can be found, for example, in \cite{fachechi2025multichannel}, where the model remains based on a modular Hebbian architecture, and in \cite{Jean2024}, which employs a Bayesian approach to achieve compelling performance.
}

\setcounter{equation}{0}
\renewcommand\theequation{G.\arabic{equation}}

\section{A performance-driven revision} \label{sec:NewModel}

The analysis carried on in this manuscript showed that an assembly of interacting Hopfield networks is able to accomplish tasks that are not achievable by a single Hopfield network. However, since the preliminary results presented in Sec.~\ref{sec:HLN}, one could realize that the simplest model we use to inspect that {\em more is different} is probably not the optimal one if specifically interested in pattern disentanglement as more complex models may perform better; indeed, our purpose is the investigation of non-trivial phenomena emerging from the interaction of networks, rather than specifically pattern disentanglement, see \cite{Jean2024}. In fact, our target configuration is not a ground state for the model and, as $\beta \to \infty$, the system would remain stuck in the input configuration. We recognize that the intra-layer interactions work properly by favoring the alignment of each layer to patterns, on the other hand, the inter-layer interactions, which should inhibit the retrieval of the same pattern by different layers, tend to favor the staggered configuration instead of the target configuration. This flaw can be fixed by revising the coupling between different layers. Indeed, this term explicitly breaks the layer-wise spin-flip symmetry of our model and stabilizes the state $\bm{\sigma}^{(1,1,1')}=(\bm{\xi}^1,\bm{\xi}^1, -\bm{\xi}^1)$, which is among the states that most significantly hinder the network's disentanglement task.
%Therefore, in order to restore gauge invariance to these terms inhibit thermalization into states like $\bm\sigma^{(1,1,1')}$, a modification is necessary.
%
A modified cost function reads as:
\begin{equation} \label{eq:HamHam_quad}
\begin{array}{lll}
      \widetilde{\mathcal{H}}(\boldsymbol \sigma ; \lambda, H, \boldsymbol \xi, \boldsymbol h ) = -N\SOMMA{\mu=1}{K} \sum_{a=1}^L (m^a_{\mu})^2 - H \SOMMA{i=1}{N}\sum_{a=1}^L h^{a}_i\sigma_i^a + N\lambda\sum_{\substack{a,b=1 \\ a\neq b}}^L \left(\SOMMA{\mu=1}{K} m^a_\mu m^b_\mu  \right)^2
\end{array}
\end{equation}
%%%%%%%%%%%
and it differs from the original one \eqref{eq:HamHam} only in the last contribution in the right-hand side of \eqref{eq:HamHam_quad}, which now features a quadratic sum over the heterogeneous product of magnetizations, rather than a linear one.
This modification has two advantages: first, in the absence of an external field (i.e., $H = 0$), it makes the cost function invariant under layer-wise spin-flip, further, it inhibits the relaxation towards states like $\bm \sigma^{(1,1,1’)}$, making, as we will see, the disentanglement task more robust and stable even at very low noise.

\begin{figure}[tb]
    \centering
    \includegraphics[width=13cm]{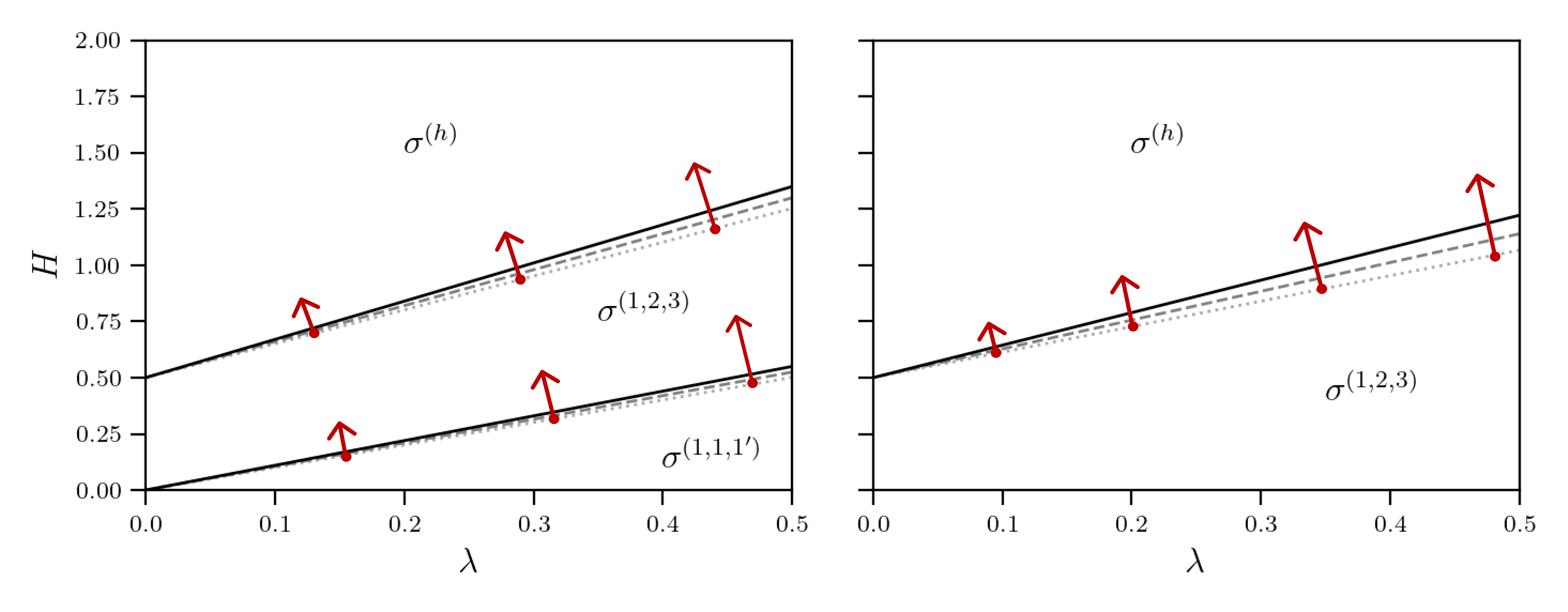}
    \caption{We evaluate $\mathcal H (\boldsymbol \sigma; \lambda, H, \boldsymbol \xi, \boldsymbol h)/N$ (left panel) and $\widetilde{\mathcal H} (\boldsymbol \sigma; \lambda, H, \boldsymbol \xi, \boldsymbol h)/N$ at the configurations $\boldsymbol \sigma^{(1,2,3)}$, $\boldsymbol \sigma^{(1,1,1)}$, $\boldsymbol \sigma^{(1,1,1')}$, and $\boldsymbol \sigma^{(h)}$ according to eqs.~\eqref{eq:HHH0},\eqref{eq:HHH11_1}, \eqref{eq:HHH1} and \eqref{eq:domaron}-\eqref{eq:trimaron}, and we keep track of the configurations displaying the lower energy as the parameters $H$ and $\lambda$ are varied. For the first model the region where the target configuration is energetically favoured is the one corresponding to relatively large values of $\lambda$ and relatively small values of $H$, while for the second model that region encompasses the whole region below the curve $H=1/2 + 2 \lambda (3/4 + \gamma)^2$. Different values of $\gamma$ are also considered: $\gamma = 0.1$ (solid line), $\gamma = 0.05$ (dashed line), and $\gamma = 0.005$ (dotted line): the arrows point in the direction of increasing $\gamma$.}
    \label{fig:diagdiag}
\end{figure}

An easy and intuitive way to see that is by looking at the energies associated to the configurations treated in Sec.~\ref{sec:HLN}, that now read as 
\begin{eqnarray} 
\label{eq:domaron}
\frac{\widetilde{ \mathcal H}(\boldsymbol \sigma^{(1,2,3)})}{N} \underset{c.l.t.}{\sim} -3(1+\gamma) - \frac{3}{2}H+x\frac{\tilde{\mathcal{C}}^{(1,2,3)}}{\sqrt{N}}, 
\\
 \frac{\widetilde{\mathcal H} (\boldsymbol \sigma^{(1,1,1)})}{N} \underset{c.l.t.}{\sim}-3(1+\gamma) +3\lambda(1+\gamma)^2-\dfrac{3}{2}H+x\frac{\tilde{\mathcal{C}}^{(1,1,1)}}{\sqrt{N}},
 \\
 \frac{ \widetilde{\mathcal H} (\boldsymbol \sigma^{(1,1,1')})}{N} \underset{c.l.t.}{\sim}-3(1+\gamma) +3\lambda(1+\gamma)^2-\dfrac{1}{2}H+x\frac{\tilde{\mathcal{C}}^{(1,1,1')}}{\sqrt{N}}
 \\
 \label{eq:trimaron}
 \frac{ \widetilde{\mathcal H} (\boldsymbol \sigma^{(h)})}{N} \underset{c.l.t.}{\sim} -3\left(\dfrac{3}{4}+\gamma\right)+3\lambda\left(\dfrac{3}{4}+\gamma\right)^2-3H+x\frac{\tilde{\mathcal{C}}^{(h)}}{\sqrt{N}}.
\end{eqnarray}
%%%%%%%%%%%%%
By comparison with eqs.~\eqref{eq:HHH0}, \eqref{eq:HHH}, \eqref{eq:HHH11_1}, and \eqref{eq:HHH1}, we see that $\widetilde{ \mathcal H(}\boldsymbol \sigma^{(1,2,3)})$ is asymptotically the same as ${ \mathcal H(\boldsymbol \sigma^{(1,2,3)})}$, moreover, now $\lambda$ has a stronger effect in making the configuration $\boldsymbol \sigma^{(1,1,1)}$ unstable and its influence on $\boldsymbol \sigma^{(1,1,1')}$ shifts from positive to negative; as for $\boldsymbol \sigma^{(h)}$, this state is slightly favored in the current setting, especially for low loads.
As a result, here, for $H$ relatively small, $\boldsymbol \sigma^{(1,2,3)}$ is always prevailing over $\boldsymbol \sigma^{(1,1,1')}$, see Figure~\ref{fig:diagdiag}.

\begin{figure}[tb]
    \centering
    \includegraphics[width=14cm]{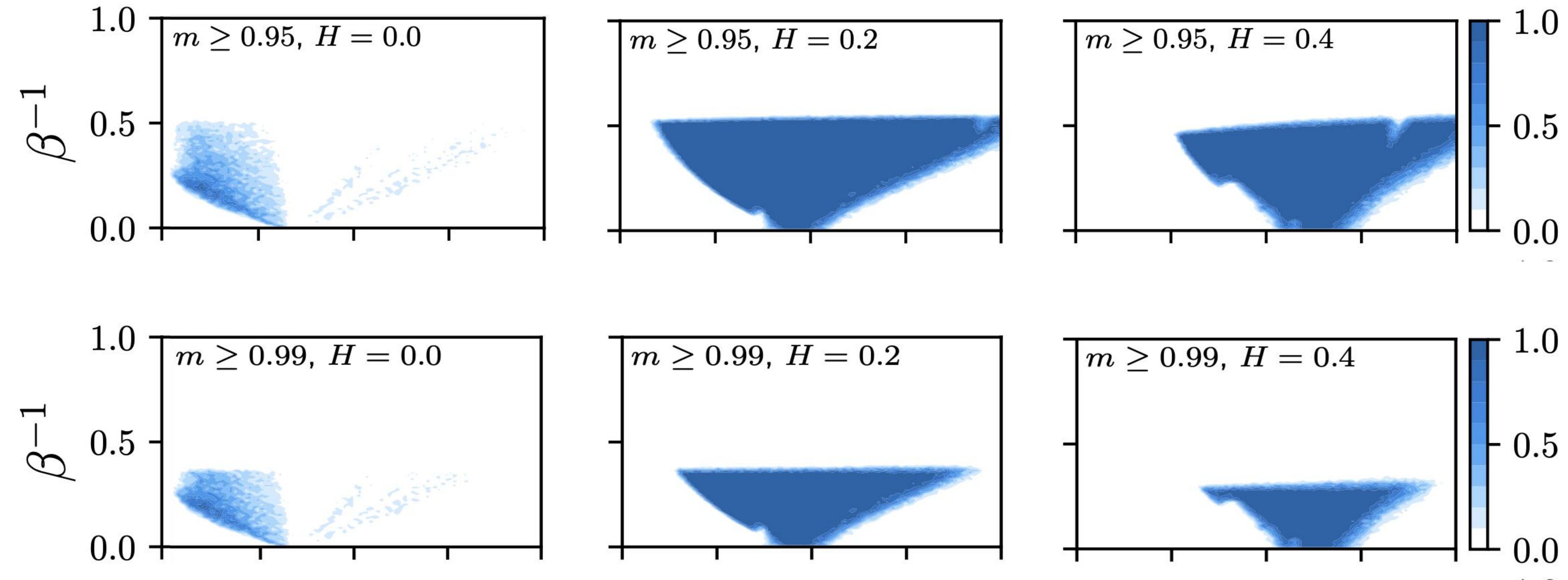}
    \caption{
    We consider the system described by the revised cost function \eqref{eq:HamHam_quad} and we simulate its evolution starting from the configuration $\bm{\sigma}^{(h)}$ and iteratively applying the noisy dynamics \eqref{eq:evolv},  up to convergence to equilibrium. Analogously with what done in Figure. \ref{fig:allyoucan} - \ref{fig:spurious-res}, we set $N=5000$ and $K=50$ and we repeated the MC simulation $50$ times for each sampled point of the ($\lambda$, $\beta^{-1}$) plane and for various values of the external field $H = 0.0$ (first column), $H = 0.2$ (second column), $H = 0.4$ (third column); next, the magnetizations of the three layers versus the patterns $\bm\xi^1$, $\bm\xi^2$, $\bm\xi^3$, are evaluated and, if each of the three patterns is retrieved with a quality at least equal to the given threshold, the simulation is considered as successful. The accuracy, represented by the color map, is then evaluated over the sample of $50$ trials. Finally, notice that, unlike Figure.~\ref{fig:allyoucan}-\ref{fig:spurious-res} here we plotted data versus $\beta^{-1}$ to highlight that the system is able to accomplish the task even in the noiseless case $\beta^{-1} \to 0$.}
    \label{fig:acc_quad}
\end{figure}

Finally, MC simulations analogous to those presented in Sec.~\ref{sec:MC} have been run for the system described by the cost function \eqref{eq:HamHam_quad} and for various parameter settings.
The results, presented in Figure~\ref{fig:acc_quad}, show that the region where the spurious-state disentanglement occurs successfully is no longer vanishing in the zero-temperature limit. Furthermore, when the temperature increases (e.g., at $\beta = 2$), the region of high accuracy performance is significantly enlarged compared to the results obtained with the cost function \eqref{eq:hamiltonian} and presented in Figure~\ref{fig:spurious-res}. 
The robustness of these results is checked in Figure~\ref{fig:digiti}, where we executed a numerical test with a nonrandom data set, where the patterns represent digits and their mixture (see the left-most panels in the figure) is used as input for a three-layer network where neurons interact according to \eqref{eq:HamHam_quad}. 
%The network is able to disentangle the mixture in a wide range of parameters capacity is presented in Figure~\ref{fig:digiti}.

\setcounter{equation}{0}

\renewcommand\theequation{H.\arabic{equation}}

\section{\red{Insight into pattern disentanglement}} \label{sec:Explanation}

\begin{figure}[tb]
    \centering
    \includegraphics[width=14cm]{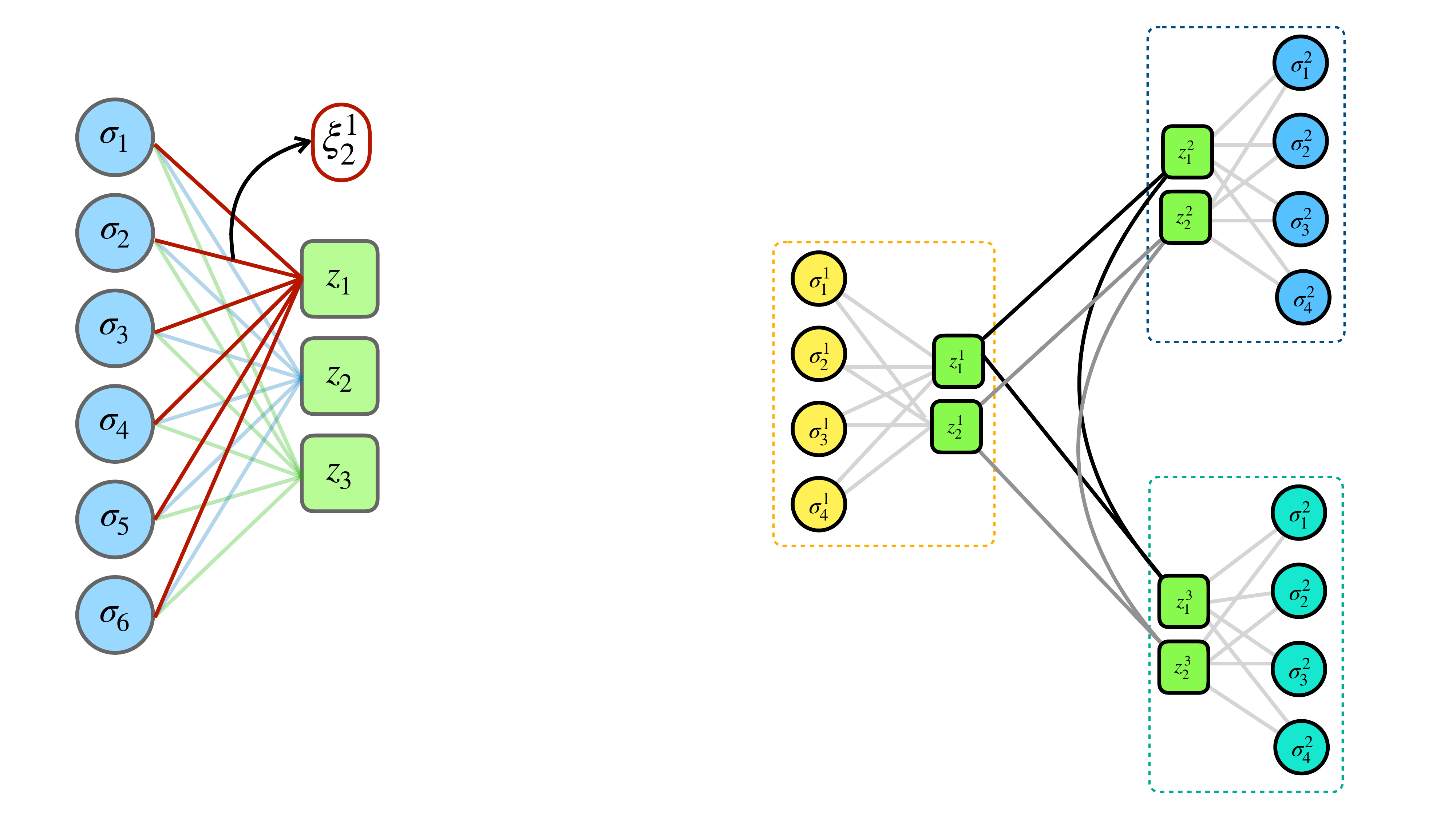}
    \caption{
    \red{This figure shows the structure of the RBM equivalent to a single HNN (left panel) and a chain of RBMs equivalent to the modular Hebbian network of Hopfield networks (right panel).  The former corresponds to a network made of $N=6$ neurons and $K=3$ patterns, the latter corresponds to a network made of $N=4$ neurons in each of the three layers and a dataset made of $K=2$ patterns: note that, in both the scenarios, the amount of hidden neurons matches the amount of stored patterns, such that the hot-vector coding (to preserve a Machine Learning jargon) -or the grandmother cell setting (to prefer a Neuroscience vocabulary)- is naturally assumed: we deepened this duality between heteroassociative Hebbian networks and generalized RBMs in  \cite{AABCR-2024}.}}
    \label{fig:duality}
\end{figure}

\red{The main reward in having a theory rather than empirical algorithms is probably the explainability it may offer and, in this appendix, by relying upon the theory exposed in the main text, we try to explain why the pattern disentanglement mechanism provided by these Hebbian networks can be a rationale also for understanding pattern disentanglement by deep learning scaffolds build of by chains of restricted Boltzmann machines \cite{Cacao20,DRL2024}. The key ingredient that we need is bridging Hopfield neural networks (HNN) and restricted Boltzmann machines (RBM) \cite{AABF-NN2020}, leveraging the grandmother cell setting as we briefly recall. Retaining a dataset made of random patterns $\{\boldsymbol \xi^{\mu}\}_{\mu=1, ..., K}$, the cost function of a single Hopfield network built of by $N$ binary neurons $\sigma_i, \ \ i \in (1,...,N)$ reads as $\mathcal{H}^{\textrm{HNN}}(\boldsymbol{\sigma};\boldsymbol{\xi}) = -(1/2N)\sum_{i<j}\sum_{\mu}^K \xi_i^{\mu}\xi_j^{\mu}\sigma_i \sigma_j$ and its partition function $\mathcal{Z}_N^{\textrm{HNN}}(\beta; \boldsymbol{\xi})$ can be written as}
\begin{eqnarray}\label{bare-duality}
\mathcal{Z}^{\textrm{HNN}}_N(\beta; \boldsymbol{\xi}) &=& \sum_{\{\bm \sigma\}} e^{-\beta \mathcal{H}^{\textrm{HNN}}(\boldsymbol{\sigma};\boldsymbol{\xi})} =\sum_{\sigma}^{2^N} e^{\frac{\beta}{2N} \sum_{i<j} \sum_{\mu}^K \xi_i^{\mu}\xi_j^{\mu}\sigma_i \sigma_j} \\
\label{eq:h1}
&=& \sum_{\sigma}^{2^N} \int \prod_{\mu}^K dz_{\mu} e^{-\sum_{\mu}^K \frac{ \beta z_{\mu}^2}{2}} e^{\frac{\beta}{\sqrt{N}} \sum_{i,\mu}^{N,K} \xi_i^{\mu}\sigma_i z_{\mu}}=\mathcal{Z}^{\textrm{RBM}}_N(\beta; \boldsymbol{\xi}).
\end{eqnarray}
\red{where, in the second line, we used the Gaussian integration to obtain the integral representation of the partition function of the Hopfield model. This gives rise to three essential observations (see also     Figure \ref{fig:duality}, left panel):}
%%%
\red{
\begin{itemize}
\item The exponent in the second contribution at the l.h.s. of eq.~\eqref{eq:h1} reads $\mathcal{H}^{\textrm{RBM}}(\boldsymbol{\sigma},\boldsymbol{z};\boldsymbol{\xi})= -\frac{1}{\sqrt{N}} \sum_{i,\mu}^{N,K} \xi_i^{\mu}\sigma_i z_{\mu}$ that is nothing but the cost function of a RBM equipped with a visible layer built of by the $N$ binary neurons $\{\sigma_i\}_{i=1,...,N}$ and a hidden layer built of by $K$ real-valued neurons $\{z_{\mu}\}_{\mu=1,...,K}$ displaying a Gaussian prior.
\item The pattern entries $\xi_i^{\mu}$ in the HNN play as the weights connecting the visible neuron $\sigma_i$ to the hidden neuron $z_{\mu}$ in the dual RBM.
\item The dual RBM features exactly $K$ hidden neurons, one per pattern, such that when the visible layer is inputted with a (possibly noisy) pattern, say $\boldsymbol \xi^1$ -- namely when this input is provided to the HNN -- the corresponding hidden neuron $z_1$ gets active -- mirroring the retrieval of the first pattern by the HNN -- while the other hidden neurons remain silent. With this hot vector coding, a hidden neuron can therefore be interpreted as a grandmother cell in Neuroscience, that is a highly selective hidden neuron responding solely to a specific pattern\footnote{\red{This can be intuitively understood by noting that the cost function of the RBM can be recast in terms of the Mattis magnetization as $\mathcal{H}^{\textrm{RBM}}(\boldsymbol{\sigma},\boldsymbol{z};\boldsymbol{\xi}) = - \sqrt{N} \sum_{\mu} m_{\mu} z_{\mu}$, such that, when $m_{\mu} \sim 1$ -- namely the HNN has retrieved the $mu$-th pattern -- the field experienced by the neuron $z_{\mu}$ gets large, thus forcing the neuron to get active, while the fields experienced by the remaining hidden neurons remain negligible.}}.
\end{itemize}}
%%%
%
%%%
\red{
Clearly, if the HNN is inputted with a mixture of patterns, it gets stuck into a spurious state and, accordingly, the corresponding dual RBM shows multiple mildly active hidden neurons. However, by generalizing the above picture of the integral representation of Hebbian networks in terms of RBMs, we can explain why this does not happen in hetero-associative networks and why their architecture actually corresponds to networks of RBMs reminiscent of those used in deep learning \cite{Cacao20}. We discuss solely the simplest scenario of $L=3$ and start by resuming cost function}
%%%
\begin{equation} \label{eq:Ham_AppH}
\mathcal{H}(\boldsymbol{\sigma};\lambda, H,\boldsymbol{\xi},\boldsymbol{h}) = - N \sum_{\mu=1}^K \sum_{a=1}^3 (m_{\mu}^a)^2 -H \sum_{i=1}^N \sum_{a=1}^3 h_i^a \sigma_i^a + N\frac{\lambda}{2}\sum_{\substack{a,b=1 \\ a \neq b}}^{3} m_{\mu}^a m_{\mu}^b.
\end{equation}
%%%%
\red{Its partition function along with its integral representation can be written as}
\begin{equation}
\begin{array}{lll}
     \mathcal{Z}^{\textrm{HNN}}_N(\beta,H=0,\boldsymbol g, \boldsymbol \xi, \boldsymbol h) &= \SOMMA{\{\bm\sigma^{1}\}}{}\cdots\SOMMA{\{\bm\sigma^{L}\}}{}e^{\frac{\beta}{2N} \sum\limits_{a,b=1}^{L,L}g_{ab}\sum\limits_{\mu,i,j=1}^{K,N,N} \xi_i^\mu\xi_j^\mu  \sigma_i^a\sigma_j^b}
     \\\\
     &= \SOMMA{\{\bm\sigma^{1}\}}{}\cdots\SOMMA{\{\bm\sigma^{L}\}}{}\displaystyle\int \prod\limits_{\mu,a,b=1}^{K,L,L}\dfrac{dz_{\mu}^adz_{\mu}^b}{2\pi}e^{-\frac{\beta}{2}\sum\limits_{\mu,a,b=1}^{K,L,L}z_{\mu}^a(\bm g^{-1})_{ab} z_{\mu}^b+\frac{\beta}{\sqrt{N}} \sum\limits_{\mu,a,i=1}^{K,L,N} \xi_i^\mu \sigma_i^a z_{\mu}^a}\\ \\
     &=\mathcal{Z}_N^{\textrm{RBM}}(\beta,H=0,\boldsymbol g, \boldsymbol \xi, \boldsymbol h)
\end{array}
\end{equation}
\red{We can see that the first contribution in \eqref{eq:Ham_AppH} (i.e., $-N \sum_{\mu} \sum_a (m_{\mu}^a)^2$) has an integral representation in terms of three independent RBMs and the third contribution (i.e., $+ N\frac{\lambda}{2}\sum_{a \neq b} m_{\mu}^a m_{\mu}^b$) yields a repulsive (note the reversed sign w.r.t. the first term) interaction  among their hidden layers, see Figure \ref{fig:duality}, right panel; the second contribution provides the external input to their visible layers and it is not affected by the integral representation. Consequently, when a mixture of patterns is presented to the visible layers of these machines, each RBM attempts to retrieve a specific pattern by activating its corresponding grandmother neuron (due to the first term), while the last term ensures that they do not retrieve the same pattern. This mechanism promotes the disentanglement of input mixtures and prevents the network from becoming trapped in spurious states. 
\newline
In the dual representation of this network of Hebbian networks, the architecture connecting the various RBMs -each specialized in the retrieval of a given pattern- strongly resembles that of a deep learning scaffold built up to RMBs once trained to accomplish the disentanglement task  \cite{Cacao20}.
}

\newpage

\bibliographystyle{unsrt}
\bibliography{NeuralNetworks}

\end{document}